\documentclass[runningheads]{llncs}

\usepackage[utf8]{inputenc}
\usepackage{listings}
\usepackage{amsmath}
\usepackage{amssymb} 
\usepackage{xspace}
\usepackage{tikz} 
\usepackage{tikz-qtree}
\usepackage{graphics} 
\usepackage{array}
\usepackage{multirow} 
\usepackage{float} 
\usepackage{leftidx}
\usepackage{enumerate}
\usepackage{booktabs}
\usepackage{comment}
\usepackage{shuffle}

\let\oldstar\star

\DeclareMathOperator{\REV}{R}
\DeclareMathOperator{\rank}{\mathsf{rank}}
\DeclareMathOperator{\width}{\mathsf{width}}
\DeclareMathOperator{\dsc}{\mathsf{sc}}
\DeclareMathOperator{\nsc}{\mathsf{nsc}}
\DeclareMathOperator{\bm}{\mathsf{B}}
\DeclareMathOperator{\ind}{\mathsf{ind}}
\DeclareMathOperator{\pad}{\mathsf{pad}}
\DeclareMathOperator{\bit}{\mathsf{last}}
\DeclareMathOperator{\rotation}{\mathsf{R}}

\newcommand{\Mod}[1]{\ (\mathrm{mod}\ #1)}
\newcommand{\naturals}{\mathbb{N}}

\newcommand{\alphabet}{\Sigma}
\newcommand{\letter}{\sigma}
\newcommand{\reversal}[1]{#1^{\REV}}
\newcommand{\extended}[1]{#1_{st}}

\newcommand{\plus}[1]{#1^+}
\newcommand{\derivative}[1]{#1^{-1}}
\newcommand{\aut}[1]{\mathcal{#1}}
\newcommand{\auttuple}[5]{\left\langle#1,#2,#3,#4,#5\right\rangle}
\newcommand{\lang}[1]{\mathcal{L}(#1)}
\newcommand{\stateleftlang}[2]{\overleftarrow{\mathcal{L}}_{#2}(#1)}
\newcommand{\staterightlang}[2]{\mathcal{L}_{#2}(#1)}
\newcommand{\determine}[1]{\mathcal{D}(#1)}

\newcommand{\bmsegment}[2]{\mathsf{s}^{#1}_{#2}}

\newcommand{\bmset}[1]{\mathcal{B}_{#1}}
\newcommand{\bmsetl}[2]{\mathcal{B}(#1)_{#2}}
\newcommand{\cover}[1]{\mathcal{C}_{#1}}
\newcommand{\bin}[1]{{#1}_{[2]}}
\newcommand{\dfawitness}{MAX_{\ell}}
\newcommand{\dfawitnessinst}[1]{MAX_{#1}}
\newcommand{\maxr}{r_{\ell}}

\newcommand{\m}{t_\ell}
\renewcommand{\complement}[1]{\overline{#1}}
\renewcommand{\star}[1]{#1^\oldstar}
\renewcommand{\epsilon}{\varepsilon}
\newcommand{\dfa}{DFA\xspace}
\newcommand{\dfas}{DFAs\xspace}
\newcommand{\nfa}{NFA\xspace}
\newcommand{\nfas}{NFAs\xspace}
\newcommand{\cset}[1]{\{\,#1\,\}}

\usetikzlibrary{cd,shapes,shapes.geometric,arrows,fit,calc,positioning,automata,decorations.pathmorphing,shadows} 
\tikzset{sstate/.style={state, inner sep=2pt, minimum size=7pt, circular drop shadow, fill=white}}
\tikzset{phantom/.style={state, white, inner sep=2pt, minimum size=7pt, fill=white}}

\lstnewenvironment{algorithm}[1][] {
    \lstset{
		mathescape=true,
       	frame=tB,
       	numbers=left, 
       	numberstyle=\scriptsize,
       	basicstyle=\small, 
       	keywordstyle=\color{black}\bfseries,
       	keywords={,input, output, return, in, if, else, for, while, begin, end, then, def, assert, let, do, foreach, from, to,} 
        numbers=left,
        xleftmargin=.04\textwidth,
        #1
	}
}{}

\title{On the Representation and State Complexity of Block Languages}
\titlerunning{On Block Languages}

\author{
	Guilherme Duarte\orcidID{0000-0002-4119-0694}\inst{1} \and
	Nelma Moreira\orcidID{0000-0003-0861-0105}\inst{1}\and
	Luca Prigioniero\orcidID{0000-0001-7163-4965}\inst{2}\and 
	Rogério Reis\orcidID{0000-0001-9668-0917}\inst{1}
	\thanks{This work was partially supported by CMUP through the FCT projects UIDB/00144/2020 and UIDP/00144/2020.	
	}
}
\institute{
	CMUP \& DCC, Faculdade de Ci\^encias da Universidade do Porto\\
	Rua do Campo Alegre, 4169-007 Porto, Portugal\\
	\email{\{guilherme.duarte,nelma.moreira,rogerio.reis\}@fc.up.pt}
	\and
	Department of Computer Science, Loughborough University\\ Epinal Way, Loughborough LE11 3TU, United Kingdom\\
	\email{L.Prigioniero@lboro.ac.uk}
}
\authorrunning{
	G.~Duarte\and
	N.~Moreira\and
	L.~Prigioniero\and
	R.~Reis
}

\begin{document}

\maketitle
	
\begin{abstract}
In this paper, we consider block languages, namely sets of words having the same length, and we propose a new representation for these languages. In particular, given an alphabet of size $k$ and a length $\ell$, a block language can be represented by a bitmap of length $k^\ell$, where each bit indicates whether the corresponding word, according to the lexicographical order, belongs, or not, to the language (bit equal to 1 or 0, respectively). First, we show how to convert bitmaps into deterministic and nondeterministic finite automata, and we prove that the machines are minimal. Then, we give an analysis of the maximum number of states sufficient to accept every block language in the deterministic and nondeterministic case. Finally, we study the deterministic and nondeterministic state complexity of several operations on these languages. Being a subclass of finite languages, the upper bounds of operational state complexity known for finite languages apply for block languages as well. However, in several cases, smaller values were found. 
\end{abstract}

\section{Introduction} \label{sec:introduction}
In the area of formal languages and automata theory, the class of regular languages is one of the most investigated. Classical recognizers for this class are finite automata, in both deterministic and nondeterministic variants. The capabilities of these machines to represent languages in a more or less succinct way have been widely studied in the area of \emph{descriptional complexity}. In this context, the \emph{size} of a model is measured in terms of number of symbols used to write down its description. In the specific case of finite automata, the number of states is often considered as a measure of complexity. In this area, the minimality of finite automata has been also studied. For example, it is well known that, given a language, the deterministic finite automaton of minimal size accepting it is unique (up to isomorphisms), and there exist efficient algorithms for the minimization of these machines~\cite{almeida12:_finit_autom_minim_algor}. The situation in the nondeterministic case is more challenging as minimal nondeterministic finite automata are not necessarily unique. Furthermore, given an integer~$n$ and a language~$L$, deciding whether there is a nondeterministic finite automaton with less than~$n$ states accepting~$L$ is a \mbox{PSPACE-hard} problem~\cite{stockmeyer73:_word_probl_requir_expon_time}.

The \emph{deterministic} (\emph{nondeterministic}) \emph{state complexity} of a regular language $L$ is the number of states of its minimal complete deterministic (nondeterministic, resp.) finite automaton. Here, we are also interested in operational complexity, that is the size of the model accepting a language resulting from an operation performed on one or more languages. In particular, the \emph{state complexity of an operation} (or \emph{operational state complexity}) on regular languages is the worst-case state complexity of a language resulting from the operation, considered as a function of the state complexities of the operands.

In this paper we consider finite languages where all words have the same length, which are called \emph{homogeneous} or \emph{block} languages. Their investigation is mainly motivated by their applications to several contexts such as code theory~\cite{DudzinskiK12,KMR:2018} and image processing~\cite{KarhumakiK21,KarhOkho:2O14}. A typical problem in code theory is the construction of (maximal) block languages (codes) capable of detecting and correcting errors. Several properties of block codes using automata theory have also been studied, e.g.~\cite{ShankarDDR03}. On the other hand, an image can be represented  by a set of words of a same length  (pixels). Then, automata can be used to generate, compress, and manipulate images. 

As a subclass of finite languages, block languages inherit some properties known for that class. For instance, the minimization of deterministic finite automata can be done in linear time in the case of finite (and hence also block) languages~\cite{revuz92}. Due to the fact that all words have the same length, there are some gains in terms of descriptional complexity. It is known that the elimination of nondeterminism from an $m$-state nondeterministic finite automaton for a block language costs $2^{\Theta(\sqrt{m})}$ in size~\cite{KarhOkho:2O14}, which is smaller than the general case, for which the cost in size is $2^{\Theta(m)}$~\cite{meyer71:_econom_of_descr_by_autom,salomaa97}. The maximum number of states of minimal deterministic finite automata for finite and block languages were studied by Câmpeanu and Ho~\cite{campeanu04}, and Hanssen and Liu determined the number of block languages that attain the maximum state complexity~\cite{Kjos-HanssenL19}. Minimal deterministic finite automata for finite languages were enumerated by Almeida et al.~\cite{almeida08}. Asymptotic estimates and exact formulae for the number of $m$-state minimal deterministic finite automata accepting finite languages over alphabets of size $k$ were obtained by J. Priez~\cite{priez15} and by Price et al.~\cite{PriceFW21}.

Here we propose a new representation for block languages. In particular, given an alphabet of size $k$ and a length $\ell$, each block language can be represented by a binary string of length $k^\ell$, also called \emph{bitmap}, in which each \emph{bit} indicates whether the correspondent word, according to the lexicographical order, belongs to the language (bit equal to $1$) or not (bit equal to $0$). Then, we show how to convert bitmaps into deterministic and nondeterministic finite automata, respectively, such that the devices yielded by these conversions have minimal size. While the conversion to deterministic finite automata can be done in polynomial time in the size of the bitmap, we prove that the transformation in the nondeterministic case is \textsc{NP-complete}.

Moreover, we also use bitmaps for studying the state complexity of operations on block languages. Due to the distinguishing property of the words sharing the same length, we study standard Boolean binary operations over block languages with the same length, as well as the block complement operation (i.e., $\alphabet^\ell\setminus L$, for a block language $L$ over an alphabet $\alphabet$ and block length $\ell>0$). Nonetheless, we also consider operations such as concatenation, Kleene star, and Kleene plus, which are not closed for the class of block languages of a given length, as well as specific operations on block languages such as word removal and addition. Table~\ref{tab:cblock} summarize those results.

The paper  extends and enhances  the results presented in~\cite{DuarteMPR:2024a,DuarteMPR:2024b} and it is organized as follows. In the next section, we fix  notation and recall  some properties of finite automata for finite languages. In Section~\ref{sec:bitmap}, we introduce the bitmap representation of block languages and the conversions from bitmap to equivalent minimal automata. Section~\ref{sec:sc} analyzes the maximal  (deterministic and nondeterministic) state complexity of block languages, as well as the operational state complexities of the operations mentioned in the previous paragraph. In Section~\ref{sec:conclusions} we summarize our results.

\section{Preliminaries} \label{sec:preliminaries}
In this section we review some basic definitions about finite automata and languages and fix notation. Given two integers $i$, $j$ with $i < j$, let $[i, j]$ denote the set of integers from $i$ to $j$, including both $i$ and $j$, namely $\{i, \ldots, j\}$. Moreover, if $i$ is equal to $0$, we omit this value, thus $[j] = \{0,\ldots, j\}$. 

Given an \emph{alphabet} $\alphabet$, a \emph{word} $w$ is a sequence of symbols from $\alphabet$, and a \emph{language} $L\subseteq \star{\alphabet}$ is a set of words on $\alphabet$. The \emph{empty word} is denoted by $\varepsilon$. The \emph{length} of a word $w$ is denoted by $|w|$. We consider the alphabet $\Sigma$ ordered. Given two words $w_1,w_2$, we say that $w_1$ is \emph{lexicographically} less than   $w_2$ if, either  $|w_1|<|w_2|$, or $|w_1|=|w_2|$ and the words can be written as $w_1=x\letter_jy_1$, $w_2=x\letter_iy_2$ with $\letter_i<\letter_j$, for $\letter_i,\letter_j\in \alphabet$.

The \emph{(left) quotient} of a language $L$ by a word $w \in \star{\alphabet}$ refers to the set $w^{-1}L = \{x \in \star{\alphabet} \mid wx \in L\}$. The \emph{reversal} of a word $w = \letter_0 \cdots \letter_{n-1}$ is denoted by $\reversal{w}$ and is obtained by reversing the order of the symbols of $w$, that is $\reversal{w} = \letter_{n-1} \cdots \letter_0$. The reversal of a language $L$ is $\reversal{L}=\cset{\reversal{w}\mid w\in L}$.

A \emph{nondeterministic finite automaton} (\nfa) is a five-tuple $\aut{A}=\langle Q,\alphabet,\delta,I, F\rangle$ where~$Q$ is a finite set of states, $\alphabet$ is a finite alphabet, $I\subseteq Q$ is the set of initial states, $F \subseteq Q$ is the set of final states, and $\delta: Q \times \alphabet \to 2^Q$ is the transition function. 

As the \emph{size} of an \nfa, we consider its number of states. Sometimes we interpret the $\delta$ function as a set, where $(q_1\xrightarrow[]{\letter}q_2)\in\delta$ if $q_2\in\delta(q_1,\sigma)$, with $q_1,q_2\in Q$ and $\sigma\in\Sigma$. The transition function can be extended to words and sets of states in the natural way. Given a state $q\in Q$, the \emph{right language} of $q$ is $\staterightlang{\aut{A}}{q} = \cset{w \in \star{\alphabet} \mid \delta(q,w)\cap F \neq \emptyset}$, and the \emph{left language} is $\stateleftlang{\aut{A}}{q} = \cset{w \in \star{\alphabet} \mid q \in\delta(I,w)}$. The \emph{language accepted} by~$\aut{A}$ is $\lang{\aut{A}}=\bigcup_{q\in I}\staterightlang{\aut{A}}{q}$. An \nfa accepting a non-empty language is \emph{trim} if every state is accessible from an initial state and every state leads to a final state. An \nfa is \emph{deterministic} (\dfa) if $|I|=1$ and $|\delta(q,\letter)|\leq 1$, for all $(q,\letter) \in Q\times\alphabet$. In this case, we use $I=q_0$. We can convert an \nfa~$\aut{A}$ into an equivalent \dfa $\determine{\aut{A}}$ using the well-known \emph{subset construction}. Two states $q_1$, $q_2$ are \emph{equivalent} (or \emph{indistinguishable}) if $\staterightlang{\aut{A}}{q_1}=\staterightlang{\aut{A}}{q_2}$. A \emph{minimal} \dfa has no different equivalent states, every state is reachable and it is unique up to isomorphism. The \emph{state complexity} of a language~$L$, $\dsc(L)$, is the size of its minimal \dfa. The \emph{nondeterministic state complexity} of a language $L$, $\nsc(L)$, is defined analogously. 

A trim \nfa~$\aut{A}=\langle Q,\alphabet,\delta,I, F\rangle$ for a finite language of words of size at most $\ell$ is \emph{acyclic} and \emph{ranked}. An NFA is \emph{acyclic} if it has no cycles, meaning that there are no paths of non-zero length within the automaton that allow to return to a previously-visited state. An NFA is \emph{ranked} if the set of states $Q$ can be partitioned into $\ell+1$ disjoint sets $Q_0,\ldots,Q_\ell$, called \emph{ranks}, such that the longest word accepted from each state in $Q_i$ is of length $i$. We define the \emph{width} of a rank $i$, namely $\width(i)$, as the cardinality of the set $Q_i$, and the \emph{width} of $\aut{A}$ to be the maximal width of all ranks, i.e., $\width(\aut{A})=\max_{i\in[\ell]}|Q_i|$.  In the following, for the ease of notation, we shall denote the ranks by their indices, e.g., we refer to rank $Q_i$ as rank $i$. For each $q\in Q$, we define $\rank(q) = \max\cset{|w| \mid w\in \staterightlang{q}{\aut{A}}}$. As $\aut{A}$ is acyclic, we have $0\leq\rank(q)\leq\ell$ and that all transitions from states of rank $i$ lead only to states in ranks $j$, such that~$j<i$.

A \dfa for a finite language is also ranked but it may have a \emph{sink-state} $\Omega$ which is the only state with a self-loop and without a rank. In a trim acyclic automaton, two states $q$ and $q'$ are equivalent if they are both in the same rank, either final or not final, and their transition functions lead to equivalent states, i.e., $\delta(q,w) \in F \iff \delta(q',w)\in F$, for each word $w\in\star{\alphabet}$. An acyclic \dfa can be minimized by merging equivalent states and the resulting algorithm runs in linear time in the size of the automaton (Revuz algorithm,~\cite{revuz92,almeida08}).

\section{Bitmaps and Minimal Automata for Block Languages} \label{sec:bitmap}
Given an alphabet $\alphabet=\{\letter_0,\ldots,\letter_{k-1}\}$ of size $k>1$ and an integer $\ell>0$, a \emph{block language} $L\subseteq\alphabet^\ell$ is a set of words of length $\ell$ over $\alphabet$. The integer $\ell$ is called the \emph{block length}.

The language $L$ can be characterized by a word in $\{0,1\}^{k^\ell}$ that we call \emph{bitmap} and denote as 
$$\bm(L)= b_0\cdots  b_{k^\ell-1},$$
where $b_i$ is~$1$ if the word $w$ is in $L$ and~$0$ otherwise, and $i\in [k^\ell-1]$ is the index of $w$ in the lexicographical ordered list of all the words of $\alphabet^\ell$. In this case, we denote $i$ by $\ind(w)$. We shall denote the bitmap of a language as $\bm$ when it is clear from the context to which language the bitmap refers to. Moreover, each bitmap $\bm\in\{0,1\}^{k^\ell}$ represents a block language of length $\ell$ over a $k$-ary alphabet, thus one can use any alphabet of size $k$. We also say that $\lang{\bm}$ is the block language that $\bm$ represents.
 
\begin{example} \label{example:bitmap}
	Let $\alphabet=\{a,b\}$, $\ell=4$, and 
	$$L=\{aaaa,aaba,aabb,abab,abba,abbb,babb,bbaa,bbab,bbba\}.$$ 
	The bitmap of $L$ is $\bm=1011011100011110.$ For example, $b_0=1$ since $aaaa\in L$ and $\ind(aaaa)=0$. Also, $b_6=1$ because $\ind(abba)=6$ and $abba\in L$. On the other hand, $b_{15}=0$ as $bbbb\notin L$.
\end{example}

A bitmap $\bm \in \{0,1\}^{k^\ell}$ can be divided into segments of length $k^i$, for $i\in [\ell]$. That is, the bitmap $\bm$ can be seen as
	$$\bm = \bmsegment{i}{0}\cdots\bmsegment{i}{k^{\ell-i}-1},$$
	such that $\bmsegment{i}{j}=b_{jk^i}\cdots b_{(j+1) k^i-1}$ denotes the $j$-th segment of length $k^i$, for each $j\in[k^{\ell-i}-1]$. Whenever $i>0$, each segment can also be split into $k$ segments, so $\bmsegment{i}{j}$ is inductively defined as
\begin{align*}
	\bmsegment{i}{j}=
	\begin{cases}
		b_j, & \text{if } i = 0, \\
		\bmsegment{i-1}{jk}\cdots\bmsegment{i-1}{(j+1)k-1}, & \text{otherwise.}
	\end{cases}
\end{align*}

\begin{example}\label{example:segment}
	Recall Example~\ref{example:bitmap}, where $\bm=1011011100011110$, $k=2$ and $\ell=4$.  For $i=0$, we have  $\bmsegment{0}{0}=\bmsegment{0}{2}=\bmsegment{0}{3}=\bmsegment{0}{5}=\bmsegment{0}{6}=\bmsegment{0}{7}=\bmsegment{0}{11}=\bmsegment{0}{12}=\bmsegment{0}{13}=\bmsegment{0}{14}=1$  and $\bmsegment{0}{j}=0$, for the remaining $j$.  For $i=1$ ($\bm=10\ 11\ 01\ 11\ 00\ 01\ 11\ 10$), we have  
		$$ \bmsegment{1}{0}=\bmsegment{1}{7}=10, \, \bmsegment{1}{1}=\bmsegment{1}{3}=\bmsegment{1}{6}=11, \, \bmsegment{1}{2}=\bmsegment{1}{5}=01, \, \text{and }  \bmsegment{1}{4}=00.$$
		For $i=2$ ($\bm=1011\ 0111\ 0001\ 1110$), we have
		$$ \bmsegment{2}{0}=\bmsegment{1}{0}\bmsegment{1}{1}=1011, \, \bmsegment{2}{1}=\bmsegment{1}{2}\bmsegment{1}{3}=0111, \, \bmsegment{2}{2}=\bmsegment{1}{4}\bmsegment{1}{5}=0001, \, \text{and } \bmsegment{2}{3}=\bmsegment{1}{6}\bmsegment{1}{7}=1110.$$
                For $i=3$ ($\bm=10110111\ 00011110$), we have
                	$ \bmsegment{3}{0}=\bmsegment{2}{0}\bmsegment{2}{1}=10110111$ and $ \bmsegment{3}{1}=\bmsegment{2}{2}\bmsegment{2}{3}=00011110$.
    For $i=4$, we have $\bmsegment{4}{0}=\bm$.
\end{example}

Each segment of  length $k^i$  of a bitmap of a language $L\subseteq \alphabet^\ell$  and $i\in [\ell]$, represents a  (left) quotient of $L$ by some word $w\in  \alphabet^{\ell-i}$. Formally, we have
\begin{lemma} \label{lemma:segments}
	Let $L \subseteq\alphabet^\ell$ be a block language, $|\alphabet|=k$, $\ell >0$, and $\bm$ the bitmap of~$L$. Let $i\in [\ell]$, $j\in[k^{\ell-i}-1]$, and $w\in\alphabet^{\ell-i}$ be the word of index $j$ of length~$\ell-i$, in lexicographic order, i.e., $\ind(w)=j$. Then, $\bmsegment{i}{j}=\bm(w^{-1}L)$.
	\end{lemma}
\begin{proof}
	Let us prove by induction on $i\in[\ell]$. For $i=0$, by definition, $\bmsegment{0}{j} = b_j$. Additionally, we have that $b_j=1$ if the word $w$ is the $j$-th word in $\alphabet^\ell$ and $w\in L$. Since $|w|=\ell$, either $w^{-1}L=\{\varepsilon\}$ or $w^{-1}L=\emptyset$, according to the membership or not of $w$ in $L$.
	
	For the inductive step, since $i>0$, we have $\bmsegment{i}{j}=\bmsegment{i-1}{jk}\cdots \bmsegment{i-1}{(j+1)k-1}$. By hypothesis, $\bmsegment{{i-1}}{{jk + k_0}}=\bm(w_{k_0}^{-1}L)$, where $|w_{k_0}|=\ell-(i-1)$ and $\ind(w_{k_0})=jk+k_0$. One can observe that the words $\{w_{k_0}\}_{k_0\in[k-1]}$ are all equal on the first $\ell-i$ symbols, corresponding to the $j$-th word of size $\ell-i$, i.e., a word $w\in \alphabet^{\ell-i}$ with $\ind(w)=j$. Thus, $\bmsegment{i}{j}=\bm(w^{-1}L)$.
\end{proof}

\begin{example}\label{example:quotient}
	Consider Example~\ref{example:bitmap}, where $\bm=1011011100011110$, $\alphabet=\{a,b\}$ and $\ell=4$. For instance, we have that $\bmsegment{2}{0}=1011=\bm(\derivative{(aa)}L)=\{aa,ba,bb\}$, $\bmsegment{3}{1}=00011110=\bm(\derivative{b}L)=\{abb, baa, bab, bba\}$, and $\bmsegment{4}{0}=\bm=\bm(\derivative{\epsilon}L)=L$.
\end{example}

Given a bitmap $\bm\in\{0,1\}^{k^\ell}$, let $\bmset{i}$ be the \emph{set of segments of $\bm$ of length $k^i$}, for~$i\in[\ell]$, in which there is at least one bit different than zero. Formally,
\begin{equation*}
	\bmset{i} = \cset{s \in \{0,1\}^{k^i} \mid \exists j \in [k^{\ell-i}-1]: s = \bmsegment{i}{j} \text{ and } \bmsegment{i}{j} \neq 0^{k^i}}.
\end{equation*}

\begin{example} \label{example:bitmap-set}
	For the bitmap of Example~\ref{example:bitmap}, $\bm=1011011100011110$ with $k=2$ and $\ell=4$, we have $\bmset{0}=\{1\}$, $\bmset{1}=\{10,11,01\}$, $\bmset{2}=\{1011,0111,0001,1110\}$, $\bmset{3}=\{10110111,00011110\}$, and $\bmset{4}=\{\bm\}$.
\end{example}

The size of each $\bmset{i}$, for $i\in[\ell]$, satisfies the following.
\begin{lemma} \label{lemma:size-of-segments}
	Let $\bm\in\{0,1\}^{k^\ell}$ be a bitmap of a block language $L\subseteq\alphabet^\ell$ with $k=|\alphabet|$. Then, for each $i\in[\ell]$, $|\bmset{i}| \leq \min(k^{\ell-i}, 2^{k^i}-1)$.
\end{lemma}
\begin{proof}
Its obvious that  $|\bmset{i}| \leq 2^{k^i}-1$ and since there are at most $k^{\ell-i}$ unique segments of size $k^i$ in a bitmap of size $k^\ell$, then $|\bmset{i}|\leq k^{\ell-i}$, for $i \in [\ell]$.
\end{proof}

\paragraph{Boolean Block Languages Operations with Bitmaps.}
One can consider bitwise operations on bitmaps which correspond to set operations on block languages. Let $\bm_1=b_0\cdots b_{k^\ell-1}$, and $\bm_2=b_1'\cdots b_{k^\ell-1}'$ be two bitmaps of block languages over $\alphabet^\ell$, with $|\alphabet|=k$ and $\ell>0$. We define:
	\begin{enumerate}
		\item $\bm_1\land \bm_2 = (b_0\land b_0')\cdots(b_{k^\ell-1}\land b_{k^\ell-1}')$ and $\lang{\bm_1\land \bm_2} = \lang{\bm_1}\cap\lang{\bm_2}$;
		\item $\bm_1\lor \bm_2 = (b_0\lor b_0')\cdots(b_{k^\ell-1}\lor b_{k^\ell-1}')$ and $\lang{\bm_1\lor \bm_2} = \lang{\bm_1}\cup\lang{\bm_2}$;
		\item $\complement{\bm_1} = (\lnot b_0)\cdots(\lnot b_{k^\ell-1})$ and $\lang{\complement{\bm_1}} = \alphabet^{\ell}\setminus\lang{\bm_1}$;
	\end{enumerate}
	where~$\wedge$, $\vee$, and~$\neg$ indicate the standard Boolean operations of \emph{and}, \emph{or}, and \emph{complementation}, respectively.

\paragraph{Finite Automata for Block Languages.}  A finite automaton $\aut{A}=\auttuple{Q}{\alphabet}{\delta}{I}{F}$ for a block language $L\subseteq\alphabet^\ell$, with $\ell>0$ and $k>1$, is also acyclic and ranked as for general finite languages. Therefore, the set of states $Q$ can also be partitioned into $\ell+1$ sets $Q_0,\ldots, Q_\ell$, according to the rank of each state. However, the right language of each state $q\in Q_i$, for  $i\in[\ell]$, contains only words of length $i$, that is, $\staterightlang{\aut{A}}{q}\subseteq \alphabet^i$ is also a block language. As a consequence, transitions in $\aut{A}$ are from one rank to the previous one, i.e., if $q'\in\delta(q,\letter)$ then $q\in Q_i$ and $q'\in Q_{i-1}$, for some $i\in[1,\ell]$. Also, if $\aut{A}$ is minimal we have $|I|=|F|=1$, as every state in the respective ranks can be collapsed into a single state.

In the following sections, we shall describe how one can construct minimal finite automata, both deterministic and nondeterministic, from a bitmap representation of a block language.

\subsection{Minimal \dfas for Block Languages} \label{sec:dfa-construction}
In this section we relate the bitmap of a block language to its minimal DFA. Given a bitmap $\bm$ representing a block language $L \subseteq\alphabet^\ell$, with $\alphabet=\{\letter_1,\ldots,\letter_k\}$ and $\ell >0$, one can directly build a minimal \dfa~$\aut{A}$ for $L$. Refer to Algorithm~\ref{algorithm:min-dfa-block} for the details of the construction. Let~$Q=\bigcup_{i=0}^{\ell}\bmset{i}$ be the set of states of~$\aut{A}$. The transition function~$\delta$ maps the states in~$\bmset{i}$ into the ones in~$\bmset{i-1}$, for~$i\in[1,\ell]$. We start with the final state, which corresponds to the segment~$1 \in \bmset{0}$, as well as the sink-state corresponding to the segment $0$ (lines 3-4). Then, for each rank $i = 1,2,\ldots,\ell$, we consider every segment $s \in\bmset{i}$ as a state in rank $i$ (lines 6-7). If $s= s_0\cdots s_{k-1}$, where $|s_j| = k^{i-1}$,  we set $\delta(s, \letter_j)=s_j$, for $j\in[k-1]$ (lines 8-11). The function $segment(s,n)$ partitions the segment $s$ into smaller segments of length $n$. Note that, if the language $L$ is not empty, this construction creates exactly one initial and one final state, since $|\bmset{0}| = |\bmset{\ell}|=1$. 

\begin{algorithm}[caption={Construction of the minimal \dfa for a block language from its bitmap~$\bm$.}, label={algorithm:min-dfa-block}, captionpos=b,]
def toMinDFA($\bm$,$\alphabet$,$\ell$):
  $k\gets|\alphabet|$
  $Q\gets\{1,0\}$
  $\delta\gets\{(0\xrightarrow[]{\letter}0)\}_{\letter\in\alphabet}$
  for $i$ from $1$ to $\ell$ do
    $\bmset{i}\gets segment(\bm,k^i)$
    $Q\gets Q\cup\bmset{i}$
    foreach $s\in\bmset{i}$ do
      $S\gets segment(s,k^{i-1})$
      foreach $(\letter_j, s_j)\in\alphabet\times S$ do
        $\delta\gets\delta\cup(s\xrightarrow[]{\letter_j}s_j)$
  return $\auttuple{Q}{\alphabet}{\delta}{\bm}{1}$
\end{algorithm}

The following lemma formalizes the correctness of the construction, that is, shows that the resulting \dfa accepts the language described by the given bitmap.
\begin{lemma} \label{lemma:dfa-sound}
	Let $L\subseteq\alphabet^\ell$ be a block language with bitmap $\bm$, where $\ell >0$. Then, the \dfa~$\aut{A}$ obtained by applying Algorithm~\ref{algorithm:min-dfa-block} to $\bm$ recognizes $L$, that is, $\lang{\mathcal{A}}=L$.
\end{lemma}
\begin{proof}
	Let us show that both $\lang{\mathcal{A}}\subseteq L$ and $L\subseteq\lang{\mathcal{A}}$. To show that $\lang{\mathcal{A}}\subseteq L$, let $w\in\alphabet^\ell \setminus L$. By construction, each state of $\aut{A}$ is also a segment which, by Lemma~\ref{lemma:segments}, is the bitmap of the quotient of $L$ by some word. As $w\notin L$, $w$ can be split into two words $w=w_1w_2$ such that $w_1^{-1}L = \emptyset$. As for every word $x$, $x^{-1}\emptyset=\emptyset$, we get $w^{-1}L=({w_1w_2})^{-1}L = {w_2}^{-1}(w_1^{-1}L)= \emptyset$. The empty language corresponds to the bitmap associated with the sink-state, therefore $w\notin \lang{\mathcal{A}}$. Now, let $w\in L$. Then, $w^{-1}L =\{\varepsilon\}$, whose bitmap is the final state $1$, therefore $w\in\lang{\mathcal{A}}$.
\end{proof}

Amongst all possible equivalent DFAs accepting the language represented by the given bitmap, this construction yields the minimal one, as stated in the following lemma.
\begin{lemma} \label{lemma:dfa-minimal}
	Let $L\subseteq\alphabet^\ell$ be a block language with bitmap $\bm$, where $\ell >0$. Then, the \dfa~$\aut{A}$ obtained by applying Algorithm~\ref{algorithm:min-dfa-block} to $\bm$ is minimal.
\end{lemma}
\begin{proof}
	Let $s_1, s_2$ be two distinct states of $\aut{A}$. It can be noticed that if $s_1$ and $s_2$ do not belong to the same rank, they are distinguishable. Otherwise, if they belong to the same rank, by construction, $s_1\neq s_2$, and consequently they have distinct right languages. Therefore, $s_1$ and $s_2$ are distinguishable. 
\end{proof}

Combining the results of Lemma~\ref{lemma:dfa-sound} and Lemma~\ref{lemma:dfa-minimal}, we obtain:
\begin{theorem}\label{theorem:dfa-minimal}
	Let $L\subseteq\alphabet^\ell$ be a block language. The construction presented in Algorithm~\ref{algorithm:min-dfa-block} of a \dfa from the bitmap $\bm(L)$ yields the minimal \dfa for $L$.
\end{theorem}
Algorithm~\ref{algorithm:min-dfa-block} has a complexity of $O(\ell\cdot k^\ell)$, making it linear with respect to the size of the bitmap.

\begin{example}\label{example:bitmap-to-dfa}
	Let $L\subseteq\{a,b\}^4$ be the language of Example~\ref{example:bitmap} with bitmap $\bm=1011011100011110$. The correspondent minimal \dfa obtained by applying Algorithm~\ref{algorithm:min-dfa-block} is depicted in Figure~\ref{figure:bitmap-to-dfa}. The final state, found in rank $0$, is the state $1$, and the sink-state ($0$) is omitted, as well as all transitions from and to it. States in rank~$1$ correspond to $2$-bit segments, in this case: $10$, $11$, and $01$. In particular, we have $\delta(10,a)=\delta(01,b)=\delta(11,a)=\delta(11,b)=1$. States in rank $2$ correspond to $2^2$-bit segments: $1011$, $0111$, $0001$, and $ 1110$. And we have, for instance, $\delta(0111,a)=01$ and $\delta(0111,b)=11$. Similarly for ranks $3$ and $4$. The initial state corresponds to $\bm$. 

	\begin{figure}[ht]
		\centering
	\begin{tikzpicture}[>=stealth', shorten >=1pt, auto, node distance=1.8cm, initial text={},every node/.style={scale=0.8}]
			\def\nodexshift{1.1cm}
    		\node[initial,sstate,minimum height=.5cm]   (A)                                                     {$\bm$};
			\node[sstate,ellipse,minimum height=.5cm]   (B) [above right of=A,xshift=\nodexshift,yshift=.1cm]   {{$00011110$}};
			\node[sstate,ellipse,minimum height=.5cm]   (C) [below right of=A,xshift=\nodexshift,yshift=-.1cm]  {$10110111$};
			\node[sstate,ellipse,minimum height=.5cm]   (D) [above right of=B,xshift=\nodexshift]               {$0001$};
			\node[sstate,ellipse,minimum height=.5cm]   (E) [below right of=B,xshift=\nodexshift,yshift=1.1cm]  {$0111$};
			\node[sstate,ellipse,minimum height=.5cm]   (F) [below   of=E] 
			{$1110$};
			\node[sstate,ellipse,minimum height=.5cm]   (G) [below right of=C,xshift=\nodexshift]               {$1011$};
			\node[sstate,minimum height=.5cm]           (H) [below right of=D,xshift=\nodexshift,yshift=0.5cm]   {$01$};
			\node[sstate,minimum height=.5cm]           (I) [above right of=F,xshift=\nodexshift,yshift=-.5cm]  
			{$11$};
			\node[sstate,minimum height=.5cm]           (J) [above right of=G,xshift=\nodexshift]  {$10$};
			\node[sstate,accepting,minimum height=.5cm] (L) [right of=I,xshift=\nodexshift]                     {$1$};
			\def\labelpos{.35}
			\def\labelsep{.7ex}
			\path[->,near end, inner sep=\labelsep,pos=\labelpos]
				(A) edge node [below]        {$a$}   (C)
					edge node [above]        {$b$}   (B)
				(B) edge node [above]        {$a$}   (D)
					edge node [above]        {$b$}   (F)
				(C) edge node [below]        {$a$}   (G)
					edge node [below]        {$b$}   (E)   
				(D) edge node [above]        {$b$}   (H)
				(E) edge node [above,pos=.3] {$a$}   (H)
					edge node [above]        {$b$}   (I)   
				(F) edge node [above,pos=.3] {$a$}   (I)
					edge node [above,pos=.3] {$b$}   (J) 
				(G) edge node [below]        {$a$}   (J)
					edge node [above]        {$b$}   (I)
				(H) edge node [above]        {$b$}   (L)   
				(I) edge node [above,pos=.3] {$a,b$} (L)   
				(J) edge node [below] {       $a$}   (L)                   
      		;

			\path[gray]
				(G)++(-90:5ex) node (rank 2) {rank $2$}
				(rank 2 -| C)  node (rank 3) {rank $3$}
				(rank 2 -| A)  node (rank 4) {rank $4$}
				(rank 2 -| J)  node (rank 1) {rank $1$}
				(rank 2 -| L)  node (rank 0) {rank $0$}
			;
			\foreach \r [remember=\r as \lastr (initially 0)] in {1,...,4} {
				\draw[dashed,gray] ($(rank \r)!.5!(rank \lastr)$)+(0,5.6cm) -- ($(rank \r.south)!.5!(rank \lastr.south)$);
			}
		\end{tikzpicture}
		\caption{The minimal \dfa accepting the language of Examples~\ref{example:bitmap}, with $\bm=1011011100011110$.}
		\label{figure:bitmap-to-dfa}
	\end{figure}
\end{example}

Conversely, from a \dfa~$\aut{A}=\auttuple{Q}{\alphabet}{\delta}{q_0}{F}$ for a block language $L\subseteq\alphabet^\ell$, with~$\ell>0$ and $k=|\alphabet|>1$, we can construct the bitmap of~$L$ even if~$\aut{A}$ is not minimal. The method is described in Algorithm~\ref{algorithm:states-to-bitmap}. For each state $q\in Q$, it computes the bitmap $s=s_0\cdots s_{k-1}$  of $\staterightlang{\aut{A}}{q}$, such that each $s_j$ is the bitmap of the right language of the state $\delta(q,\letter_j)$, i.e. $\staterightlang{\aut{A}}{\delta(q,\letter_j)}$, with $\letter_j$ as the $j$-th symbol of the alphabet, for $j\in[k-1]$ (lines 4-14). The bitmap of each right language of a state $q\in Q$ is stored in $\mathsf{M}(q)$ (line 15). If $q$ is final, then $\mathsf{M}(q)=1$ (lines 5-6). If $\delta(q, \letter_j)$ is undefined or points to the sink-state then $\mathsf{M}(q)=0^i$, where $i=\rank(q)$, as previously discussed (lines 12-13). The bitmap for the language recognized by $\aut{A}$ is then given by  the bitmap of the initial state $q_0$ (line 16). The correctness of this algorithm follows from the correctness of the Revuz Algorithm for the minimization of \dfas for finite languages~\cite{revuz92}. Recall that if two states are equivalent, they share the same right language and, consequently, the same bitmap representation.

\begin{algorithm}[caption={Construction of the bitmap representation of a language recognized by a~\dfa.}, label={algorithm:states-to-bitmap}, captionpos=b,]
def toBitmap($\auttuple{Q}{\alphabet}{\delta}{q_0}{F}$,$\ell$):
  $\mathsf{M}\gets \emptyset$
  for $i$ from $0$ to $\ell$ do
    foreach $q\in\cset{q\in Q \mid \rank(q)=i}$ do
      if $q\in F$ then
        $\mathsf{M}(q)\gets 1$
      else
        $s\gets\epsilon$
        foreach $\letter \in \alphabet$ do
          if $\delta(q,\sigma)\in \mathsf{M}$ then
            $s'\gets \mathsf{M}(\delta(q,\sigma))$
          else
            $s'\gets 0^{|\alphabet|^i}$
          $s\gets ss'$
        $\mathsf{M}(q)\gets s$
  return $\mathsf{M}(q_0)$ 
\end{algorithm}

\subsection{Minimal \nfas for Block Languages} \label{sec:nfa-construction}
In this section, we  describe a construction of minimal \nfas from the bitmap of a block language. We also show that the problem of finding a minimal \nfa in this case is \textsc{NP-complete}. Given a bitmap $\bm$ representing a block language $L\subseteq\alphabet^\ell$, for some block length~$\ell >0$ and $|\alphabet|=k$, one can build a minimal~\nfa similarly to the construction for minimal~DFAs given in Section~\ref{sec:dfa-construction}, by iteratively finding the minimal number of states required at each rank of the \dfa. The main difference with the deterministic counterpart is that the quotients of the language can be represented by a set of states, instead of a single one.

First, let us define the notion  of  a \emph{cover} of a binary word. Let $\cover{}$ be a finite set of binary words of length $n$, that is, $\cover{}\subseteq \{0,1\}^n$, for some $n\in\naturals$. We say that $\cover{}$ is a \emph{cover for a word} $s\in \{0,1\}^n$ (or, alternatively, $s$ \emph{is covered} by $\cover{}$) if there is a subset of words in $\cover{}$ such that the bitwise disjunction of those words equals $s$. Formally, $\cover{}$ \emph{covers} $s$ if $\exists\, \mathcal S \subseteq\cover{}$ such that~$\bigvee_{c \in \mathcal S} c = s$.

We extend this definition to sets in the natural way, that is, $\cover{}$ is a \emph{cover of a finite set} of binary words $\bmset{}\subseteq \{0,1\}^n$ if every word in $\bmset{}$ is covered by $\cover{}$. Trivially, $\bmset{}$ covers itself. We say that $\cover{}$ is a \emph{minimal cover} for $\bmset{}$ if there is no smaller set that covers $\bmset{}$.
\begin{example}\label{example:covers}
	Let $\cover{}=\{1100,1010,0001\}$ and $\bmset{}=\{1100,1110,1101,1111\}$. Then, $\cover{}$ covers $\bmset{}$ because $\cover{}$ covers every word from $\bmset{}$. Namely, $1100=1100$, $1110=1100\,\lor \,1010$, $1101=1100\,\lor \,0001$, and $1111=1100\,\lor\,1010\,\lor\,0001$. One can observe that the set $\cover{}$ is a minimal cover for $\bmset{}$, but it is not unique since $\cover{}'=\{1100,0010,0001\}$ also covers $\bmset{}$ and $|\cover{}|=|\cover{}'|$.
\end{example}

Let us now describe how to obtain an \nfa for a non-empty block language $L\subseteq\alphabet^\ell$, for some $\ell\geq 0$, an alphabet $\alphabet=\{\letter_0,\ldots,\letter_{k-1}\}$, and with bitmap~$\bm$. Algorithm~\ref{algorithm:min-nfa-block} summarizes the construction. First, we define the function $\rho:\star{\{0,1\}}\to2^{\star{\{0,1\}}}$ that maps a segment into a set that covers it. The construction starts as before, with the final state $1\in\bmset{0}$ added at rank $0$, and $\rho(1)=\{1\}$, since $\bmset{0}$ is its own minimal cover (lines 3-6). Then, for each rank $i = 1,2,\ldots,\ell$, the function $minCover$ looks for the minimal set $\cover{i}$ that covers the set $\bmset{i}$, that is, the collection of segments of $\bm$ with length~$k^i$ with at least one bit set to $1$ (lines 8-9). The states in rank $i$ in the \nfa correspond to the elements of~$\cover{i}$ (line 10). Subsequently, for each segment $s\in\bmset{i}$, we set $\rho(s)=\mathcal{S}\subseteq\cover{i}$, such that $\rho(s)$ covers $s$. In the algorithm we denote, for simplicity,~$\cover{i}(s)$ as the cover of $s$ in $\cover{i}$ (lines 11-12). The transitions from rank $i$ to rank $i-1$ are determined in a way similar to the \dfa construction. For each state $c$ in rank $i$, we divide~$c$ into $c_0\cdots c_{k-1}$ using the $segment$ function, where $|c_j|=k^{i-1}$, for every $j\in[k-1]$, and set $\delta(c,\letter_j)=\rho(c_j)$, if $c_j\neq0^{k^{i-1}}$ (lines 13-16). Note that $\rho(c_j)$ must be defined or, alternatively, $c_j$ must be a segment in $\bmset{i-1}$. Therefore, we need to limit the search space of the cover $\cover{i}$, so that each word in the set is a concatenation of $k$ words from $\bmset{i-1}$ or $0^{k^{i-1}}$. Formally, $\cover{i}\subseteq (\bmset{i-1}\cup 0^{k^{i-1}})^k$, and it corresponds to one of the input of $minCover$ (line 9). Also, as seen before, $\bmset{\ell}=\{\bm\}$, so the minimal cover for this set is~$\bmset{\ell}$ itself. This implies that $\bm$ will be the single initial state at rank $\ell$ (line 17).

\begin{algorithm}[label={algorithm:min-nfa-block}, caption={Construction of a minimal NFA for a block language $L$ from its bitmap representation $\bm$.}, captionpos=b]
def toMinNFA($\bm$,$\alphabet$,$\ell$):
  $k\gets|\alphabet|$
  $Q\gets\{1\}$
  $\delta\gets\emptyset$
  $\rho\gets\emptyset$
  $\rho(1)\gets\{1\}$
  for $i$ from $1$ to $\ell$ do
    $\bmset{i}\gets segment(\bm,k^i)$
    $\cover{i}\gets minCover(\bmset{i},\bmset{i-1}\cup\{0^{k^{i-1}}\},k,i)$
    $Q\gets Q\cup\cover{i}$
    foreach $s\in\bmset{i}$ do
      $\rho(s) \gets \cover{i}(s)$
    foreach $c\in\cover{i}$ do
      $C\gets segment(c,k)$
      foreach $(\letter_j, c_j)\in\alphabet\times C: c_j \neq 0^{k^{i-1}}$ do
        $\delta\gets\delta\cup(c\xrightarrow[]{\letter_j}\rho(c_j))$
  return $\auttuple{Q}{\alphabet}{\delta}{\bm}{1}$
\end{algorithm}

The function $minCover$ takes as input a set of segments $\bmset{i}$, for some rank $i\in[\ell]$, and returns the minimal cover~$\cover{i}$ for~$\bmset{i}$. This procedure can be implemented with the help of an SMT-solver~\cite{kroening08:_decis_proced}, wherein every segment of size $k^i$ can be represented by a bit vector of the same size, and then use the binary search on the solution to determine the minimal one. Note that each $\cover{i}$, for $i\in[\ell]$, can be found in parallel since $minCover$ only requires the set of segments $\bmset{i}$ and $\bmset{i-1}$, where transitions can be added sequentially afterward.    

Now, consider the \textsc{set-basis} problem, which is defined by
\begin{itemize}
	\item \emph{Instance}: Collection of sets $B=\{B_1,\cdots,B_l\}$ and a positive integer $m$.
	\item \emph{Question}: Is there a collection of sets $C=\{C_1,\cdots,C_m\}$ such that, for each $B_i\in B$, there exists a subset of $C$ whose union equals $B_i$?
\end{itemize}

\begin{theorem}[(Stockmeyer~\cite{stockmeyer1976})]
	The \textsc{set-basis} problem is \textsc{NP-complete} by reduction to the \textsc{vertex-cover} problem.
\end{theorem}
As a consequence, we have the following result:
\begin{corollary} \label{corollary:minNFA-npcomplete}
	The construction of the minimal NFA from a bitmap representation of a block language is \textsc{NP-complete}.
\end{corollary}
\begin{proof}
	For each rank of the NFA, we seek the minimal set of segments $\cover{i}$ that covers the set $\bmset{i}$.	By Lemma~\ref{lemma:segments}, each segment corresponds to a language that is a quotient of $L$, hence, the bitwise disjunction over segments corresponds to the union over sets. This problem corresponds precisely to the \textsc{set-basis} problem. 
\end{proof}
The construction of \nfas from bitmaps described above preserves the language, as the following lemma states.
\begin{lemma} \label{lemma:nfa-sound}
	Let $L\subseteq\alphabet^\ell$ be a block language, for some $\ell >0$, with bitmap $\bm$. Then, the \nfa~$\aut{A}$ obtained  by Algorithm~\ref{algorithm:min-nfa-block} applied to $\bm$ accepts $L$, that is, $\lang{\mathcal{A}}=L$.
\end{lemma}
\begin{proof}
	Let us apply the construction given in Section~\ref{sec:dfa-construction} of the minimal \dfa from a bitmap. For $i\in[\ell]$, let $s\in\bmset{i}$ be a state of the resulting \dfa. By construction, $\rho(s)$ covers the bitmap $s$. As $\lang{s}=\bigcup_{c\in\rho(s)}\lang{c}$, one concludes that $\lang{\aut{A}}=L$. 
\end{proof}

Moreover, the \nfa given by Algorithm~\ref{algorithm:min-nfa-block} is minimal, as showed in the next lemma.
\begin{lemma} \label{lemma:nfa-minimal}
	Let $L\subseteq\alphabet^\ell$ be a block language, for some $\ell >0$, with bitmap $\bm$. Then, the \nfa~$\aut{A}$ given by Algorithm~\ref{algorithm:min-nfa-block} applied to $\bm$ is minimal.
\end{lemma}
\begin{proof}
	Let $w\in\alphabet^{\ell-i}$, for some $i\in [\ell]$. Let $\cover{i}$ be the minimal set of segments that covers the set $\bmset{i}$, and $P$ be the set of states reachable from the initial state $q_0$ of~$\aut{A}$ after reading $w$, that is, $P=\delta(q_0, w)$. Let $P_1, P_2$ be two different non-empty subsets of $P$ and let $s_1=\bigvee_{q\in P_1} q$ and $s_2=\bigvee_{q\in P_2} q$  correspond to the bitmaps of the right languages of the set of states $P_1$ and $P_2$, respectively. If $s_1=s_2$ then $P_1$ and $P_2$ cover the same bitmaps, thus, $\cover{i}\setminus P_1$ (alternatively, $\cover{i}\setminus P_2$) would also cover the set $\bmset{i}$, and so $\cover{i}$ would not be minimal. Hence, $s_1\neq s_2$.
\end{proof}

From Lemma~\ref{lemma:nfa-sound} and Lemma~\ref{lemma:nfa-minimal}, we obtain the following:
\begin{theorem}\label{theorem:min-nfa-bitmap}
	The construction of an \nfa from a bitmap of a block language $L\subseteq\alphabet^\ell$  presented in Algorithm~\ref{algorithm:min-nfa-block} results in a minimal \nfa~$\aut{A}$ such that $\lang{\mathcal{A}} = L$.
\end{theorem}

\begin{example}\label{example:bitmap-to-nfa}
	Let $L\subseteq\{a,b\}^4$ and $\bm(L)=1011011100011110$ be the language of Example~\ref{example:bitmap} and its bitmap representation, respectively. A correspondent minimal NFA obtained by applying the above construction is depicted in Figure~\ref{figure:bitmap-to-nfa}. The final state, found in rank $0$, is the state $1$. In rank $1$, we have $\bmset{1}=\{01,11,10\}$ and $\cover{1}=\{01,10\}$ is a minimal cover for $\bmset{1}$, thus only two states are needed in this rank of the NFA. For rank $2$, $\bmset{2}=\{1011,0111,0001,1110\}$ and $\cover{2}=\{1010, 0110,0001\}$, because $1011=0001\vee1010$, $0111=0110\vee0001$, and $1110=1010\vee0110$. In rank $3$ two states are needed and rank $4$ has only the initial state. In rank $3$, $\delta(10110111,b)=\{0001,0110\}$, as this set covers $0111$. Analogously, $\delta(10110111,a)=\{0001,1010\}$, as this set covers $1011$.
\end{example}

\begin{figure}[ht]
	\centering
\begin{tikzpicture}[>=stealth', shorten >=1pt, auto, node distance=1.8cm, initial text={},every node/.style={scale=0.8}]
		\def\nodexshift{1.3cm}
   		\node[initial,sstate,minimum height=.5cm]   (A)                                                     {$\bm$};
		\node[sstate,ellipse,minimum height=.5cm]   (B) [above right of=A,xshift=\nodexshift,yshift=.1cm]   { $10110111$};
		\node[sstate,ellipse,minimum height=.5cm]   (C) [below right of=A,xshift=\nodexshift,yshift=-.1cm]  {$00011110$};
		\node[sstate,ellipse,minimum height=.5cm]   (D) [above right of=B,xshift=\nodexshift]               {$0001$};
		\node[sstate,ellipse,minimum height=.5cm]   (E) [below right of=B,xshift=\nodexshift]               {$0110$};
		\node[sstate,ellipse,minimum height=.5cm]   (F) [below right of=C,xshift=\nodexshift]               {$1010$};
		\node[sstate,minimum height=.5cm]           (G) [below right of=D,xshift=\nodexshift,yshift=.1cm]  
		 {$01$};
		\node[sstate,minimum height=.5cm]           (H) [above right of=F,xshift=\nodexshift,yshift=-.1cm]  
		{$10$};
		\node[sstate,accepting,minimum height=.5cm] (I) [right of=I,xshift=\nodexshift]                     {$1$};	
		\def\labelpos{.375}
		\def\labelsep{.7ex}
		\path[->,near end,inner sep=\labelsep,pos=\labelpos]
			(A) edge node                              {$a$}   (B)
				edge node[swap]                        {$b$}   (C)
			(B) edge node[above,pos=.2]                {$a,b$} (D)
				edge node[above]                       {$b$}   (E)
				edge node[below,inner sep=2ex,pos=.12] {$a$}   (F)
			(C) edge node[above,inner sep=2ex,pos=.12] {$a$}   (D)
				edge node[below]                       {$b$}   (E)
				edge node[below]                       {$b$}   (F)   
			(D) edge node[above]                       {$b$}   (G)
			(E) edge node[above,pos=.2]                {$a$}   (G)
				edge node[below,pos=.2]                {$b$}   (H)   
			(F) edge node[above,pos=.2]                {$a,b$} (H)
			(G) edge node[pos=.2]                      {$b$}   (I)
			(H) edge node[swap,pos=.2]                 {$a$}   (I)   
      	;
      	
		\path[gray]
			(F)++(-90:5ex) node (rank 2) {rank $2$}
			(rank 2 -| C)  node (rank 3) {rank $3$}
			(rank 2 -| A)  node (rank 4) {rank $4$}
			(rank 2 -| J)  node (rank 1) {rank $1$}
			(rank 2 -| L)  node (rank 0) {rank $0$}
		;
		\foreach \r [remember=\r as \lastr (initially 0)] in {1,...,4} {
			\draw[dashed,gray] ($(rank \r)!.5!(rank \lastr)$)+(0,5.6cm) -- ($(rank \r.south)!.5!(rank \lastr.south)$);
		}
	\end{tikzpicture}
	\caption{A minimal NFA accepting the language of Example~\ref{example:bitmap}.}
	\label{figure:bitmap-to-nfa}
\end{figure}

From an \nfa~$\aut{A}=\auttuple{Q}{\alphabet}{\delta}{I}{F}$ for a block language $L\subseteq\alphabet^\ell$, with $\ell>0$ and $k=|\alphabet|>1$, one can adapt Algorithm~\ref{algorithm:states-to-bitmap} to compute the bitmap representation of~$\lang{\aut{A}}$. As before, for each state $q\in Q$, the algorithm must compute the bitmap $s=s_0\cdots s_{k-1}$ of $\staterightlang{\aut{A}}{q}$. However, each $s_j$ must now represent the bitmap of the union of the right language of the set of states $\delta(q,\letter_j)$, where $\letter_j$ is the $j$-th symbol of the alphabet, for $j\in[k-1]$. That is, $s_j$ must correspond to the bitmap of the language $\bigcup_{q'\in\delta(q,\sigma_j)}\staterightlang{\aut{A}}{q'}$. Given this observation, lines 11-15 of Algorithm~\ref{algorithm:states-to-bitmap} must be modified to compute the bitwise disjunction of the bitmaps corresponding to the right languages of each state in $\delta(q,\letter)$, for $q\in Q$ and $\letter\in\alphabet$. Finally, the bitmap of the language recognized by the automaton corresponds to the bitwise disjunction of the bitmaps representing the languages of each initial state.

\section{Block Languages State Complexity} \label{sec:sc}
In this section, we shall analyze the state complexity of block languages, namely, the maximal size of a minimal finite automaton for a block language, both for the deterministic and nondeterministic case (Section~\ref{sec:sc-dfa} and Section~\ref{sec:sc-nfa}, respectively). In Section~\ref{sec:osc}, we present results for the operational state complexity of block languages.

\subsection{Maximal Size of Minimal DFAs} \label{sec:sc-dfa}
Câmpeanu and Ho~\cite{campeanu04} showed that the number of states of a minimal \dfa accepting a block language $L\subseteq\alphabet^\ell$, over an alphabet of size $k$ and $\ell >0$, is at most $\frac{k^{\ell-r+1}-1}{k-1}+\sum_{i=0}^{r-1}(2^{k^i}-1)+1$, where $r=\min\cset{i\in[\ell]\mid k^{\ell-i}\leq  2^{k^i}-1}$. In the next result we give an estimation of the value of $r$.

\begin{theorem} \label{theorem:sc-dfa}
	Let $\ell>0$, $k>1$, and $r=\min\{i\in[\ell]\mid k^{\ell-i} \leq 2^{k^{i}}-1\}$. Then, $r=\lfloor \log_k\ell\rfloor + 1 + x$, for some $x\in\{-1,0,1\}$.
\end{theorem}
\begin{proof}
	By definition of $r$, we have that both the following inequalities hold: $k^{\ell-r}\leq2^{k^{r}}-1$ and $k^{\ell-(r-1)} > 2^{k^{r-1}}-1$. From the first inequality we obtain:
	\begin{align*}
		k^{\ell-r} \leq 2^{k^{r}}-1 & \implies  k^{\ell-r} < 2^{k^{r}} \\	
		&\implies (\ell-r)\log_2k < k^r \\
		&\implies \log_k(\ell-r) + \log_k(\log_2k) < r \\
		&\implies \log_k(\ell-r) < r \\
		&\implies \log_k\left(\ell\left(1-\frac{r}{\ell}\right)\right) < r \\
		&\implies \log_k\ell+\log_k\left (1-\frac{r}{\ell}\right) < r \implies \log_k\ell < r.
	\end{align*}
	While, from the second inequality, we get:
	\begin{align*}
		k^{\ell-r+1} > 2^{k^{r-1}}-1 & \implies  k^{\ell-r+1} \geq 2^{k^{r-1}} \\
		& \implies (\ell-r+1)\cdot\log_2k \geq k^{r-1} \\ 
		& \implies \log_k(\ell-r+1) +\log_k(\log_2k) \geq r-1 \\  
		& \implies \log_k(\ell-r+1) > r-2 \\
		& \implies \log_k\ell +\log_k\left(1+\frac{1-r}{\ell}\right) > r-2 \implies \log_k\ell > r-2.
	\end{align*}
	Moreover, $r$ is a natural number, so it is not hard to see that $r$ can be written as the floor of $\log_k\ell$ plus a small constant, as claimed.
\end{proof}

We now present a family of \emph{witness languages} recognized by minimal \dfas of maximal size, according to the bounds given in Theorem~\ref{theorem:sc-dfa}.

We denote the binary representation of a positive integer $i$ by $\bin{i}$. The~$t$-th rightmost symbol of a word~$s=\letter_0\cdots \letter_{n-1}$ of length~$n>0$ is denoted~$\bit(s,t) = \letter_{n-t}$, where~$t\in[n-1]$. Also, recall that given $w\in \alphabet^\ell$, we denote by $\ind(w)$ the index of $w$ in the lexicographical ordered list of the words of $\alphabet^{|w|}$. Let $\ell>0$, $k=2$, and $r=\min\cset{i\in[\ell]\mid k^{\ell-i}\leq  2^{k^i}-1}$ as in Theorem~\ref{theorem:sc-dfa}. In a minimal \dfa with maximal size, the rank having the largest width is either $r$ or $r-1$, depending on whether $2^{\ell-r} > 2^{2^{r-1}}-1$ holds or not, respectively. Let $r_\ell$ be that rank and $\m=\max(2^{\ell-r}, 2^{2^{r-1}}-1)$ its width, i.e, $\width(\maxr)=\m$. Then, we consider the following family of languages over $\alphabet=\{a,b\}$ and  defined for every $\ell>0$,
\begin{align*}
	\dfawitness = \{\, w_1w_2 \mid \; & w_1\in\alphabet^{\ell-\maxr},\; w_2\in\alphabet^{\maxr},\; \\
					                  & i=\ind(w_1),\; j = \ind(w_2),\; \bit(\bin{(i+1)},j)=1 \,\},
\end{align*}

Informally, these languages contain words of length~$\ell$ that can be split into words~$w_1$ and~$w_2$ of length~$\ell-\maxr$ and $\maxr$, and indices $i=\ind(w_1)$ and $j=\ind(w_2)$, respectively, such that the $j$-th rightmost symbol of $\bin{(i+1)}$ is $1$.

\begin{example} \label{example:witness-sc-dfa}
	For $\ell=5$, we have $r=\min\{i\in[5]\mid 2^{5-i}\leq 2^{2^{i}}-1\}=2$. Moreover, $\m=\max(2^{5-2}, 2^{2^{2-1}}-1)=\max(8,3)=8$, implying that $r_\ell=r$. For this configuration we have
	\begin{align*}
		\dfawitnessinst{5} =
			\{ & aaaaa, aabab, abaaa, ababa, abbba, baaaa,    \\
		       & baaba, babab, babba, bbaaa, bbaab, bbaba, bbbbb\}\text.
	\end{align*}
	For example, let $w_1 = baa$ where $i=\ind(w_1)=4$ and $\bin{(i+1)}=101$. For $j=0$ and~$j=2$, we have that $\bit(\bin{(i+1)},j)=1$, which correspond to the words $aa$ and $ba$, respectively. Thus, $\{baaaa, baaba\} \subset \dfawitnessinst{5}$.	 One can also  see that the bitmap of $\dfawitnessinst{5}$ is
	$$\bm(\dfawitnessinst{5})=10000100110000101010011011100001$$
	which corresponds to the concatenation of the  reversal of the binary representation of the first $8$ integers with $4$ bits each. Next, we explain why this is the case.
\end{example}

Let $\pad(s,j)$ be the function that concatenates leading zeros to a binary word $s$ until its length equals $j$. The bitmap of the language $\dfawitness$, for $\ell>0$, corresponds to sequences of binary representations of the first $\m$ positive integers, as follows:
\begin{lemma} \label{lemma:bitmap-witness-sc-dfa}
	Let $r$, $\maxr$, and $\m$ be defined as before for $\ell>0$ and alphabet size $k=2$. Let
	$$ P_{\m,\maxr}=\prod_{i=1}^{\m} \reversal{\pad(\bin{i}, 2^{\maxr})}.$$
	Then, the bitmap of the language $\dfawitness$ is given by
	\begin{align*}
		\bm(\dfawitness)=
		\begin{cases}
			P_{\m,\maxr},                    & \text{if } \m=2^{\ell-r}; \\
			P_{\m,\maxr}\cdot 0^{2^{\maxr}}, & \text{if } \m=2^{2^{r-1}}-1.
		\end{cases}
	\end{align*}
\end{lemma}	
\begin{proof}
	Let us show that the bitmap is correct, for either value of $\m$, by verifying both that the bitmap has length $2^\ell$, which is not trivial for the case $\m=2^{2^{r-1}}-1$, and that it represents the language $\dfawitness$.
	\begin{enumerate}
		\item $\m=2^{\ell-r}$: \label{item:bitmap-witness-case-1} \\
			In this case $\maxr=r$. Also, $|\bm(\dfawitness)|=2^{\ell-r} 2^{\maxr}=2^{\ell}$, so the bitmap has the intended length. \\
			Now, let $w_1\in\Sigma^{\ell-\maxr}$, $i=\ind(w_1)$, $w_2\in\Sigma^{\maxr}$, and $j=\ind(w_2)$. We must prove that $w_2\in w_1^{-1}\dfawitness$ if and only if the $j$-th most significant bit of $\reversal{(i+1)_{[2]}}$ is~$1$. If $w_2\in w_1^{-1}\dfawitness$ then, by definition, the $j$-th bit of the binary representation of $i+1$ is~$1$, that is, $\bit(\bin{(i+1)},j)=1$, thus the condition holds. In the other direction a similar argument applies.
		\item $\m=2^{2^{r-1}}-1$: \label{item:bitmap-witness-case-2} \\			
			Now, $r_\ell=r-1$. Let us first prove that the size of $\bm(\dfawitness)$ is $2^\ell$. From the value of $\m$, we have
			\begin{align*}
				\m \geq 2^{\ell-r} & \implies 2^{2^{r-1}}-1 \geq 2^{\ell-r} \\
				  				   & \implies 2^{2^{r-1}} \geq 2^{\ell-r} \\
								   & \implies 2^{r-1} > \ell-r.
			\end{align*}
			From the definition of $r$, we have that $\m\leq 2^{\ell-r+1}$, leading to:
			\begin{align*}
				\m \leq 2^{\ell-r+1} & \implies 2^{2^{r-1}}-1 \leq 2^{\ell-r+1} \\
				                     & \implies 2^{2^{r-1}} \leq 2^{\ell-r+1} \tag*{because $2^{2^{r-1}} > 1$} \\
				                     & \implies 2^{r-1} \leq \ell-r+1.
			\end{align*}
			These two conditions imply that \mbox{$2^{r-1}=\ell-r+1$}, so we have $|\bm(\dfawitness)| = 2^{r_\ell}2^{2^{r-1}}=2^{r_\ell+\ell-r+1}=2^{\ell}$, as desired. \\
			By~(\ref{item:bitmap-witness-case-1}) we already ensure that the bits equal to $1$ in the bitmap correspond to words that are in $\dfawitness$. Then, let us prove that the padding of $2^{r_\ell}$ zeros concatenated at the end of the bitmap correspond to words that do not belong to $\dfawitness$. Let $w_1\in\Sigma^{\ell-\maxr}$ such that $\ind(w_1)=\m=2^{2^{r-1}}-1$. In fact, the last $2^{r_\ell}$ zeros of the bitmap represent the language $w_1^{-1}\dfawitness$, thus, we need to prove that $w_1^{-1}\dfawitness = \emptyset$. By the definition of $\dfawitness$, for $\bit(\bin{(i+1)},j)=1$ to hold, $j$ must be at least $2^{r-1}$. However, we have $j=\ind(w_2)\leq 2^{r-1}-1$, for all $w_2\in\alphabet^{r_\ell}$. Thus, $w_1^{-1}\dfawitness=\emptyset$.
	\end{enumerate}
\end{proof}

\begin{example} 
	By Lemma~\ref{lemma:bitmap-witness-sc-dfa}, the bitmap of the language $\dfawitnessinst{5}$ from Example~\ref{example:witness-sc-dfa}, where $\m=8$ and $\maxr=2$, is (as already given)
	$$\bm(\dfawitnessinst{5})=\prod_{i=1}^{8} \reversal{\pad(i_{[2]}, 4)}=1000\;0100\;1100\;0010\;1010\;0110\;1110\;0001\text.$$
\end{example}

To have a \dfa of maximal size for a block language contained in $\alphabet^\ell$, for some~$\ell\geq 0$, the width of each rank $i$ must be either $2^{2^{i}}-1$, for $i\in[r-1]$, or $2^{\ell-i}$, for $i\in[r,\ell]$, from which the result from Câmpeanu and Ho was established~\cite{campeanu04}. We now prove that the minimal \dfa for $\dfawitness$ is of maximal size.

\begin{lemma} \label{lemma:witness-maxminDFA}
	Let $r$, $\maxr$, and $\m$ be defined as before for $\ell>0$ and alphabet size $k=2$. Then, the minimal \dfa accepting the language $\dfawitness$ has maximal size.
\end{lemma}
\begin{proof}
	Let $\bm$ be the bitmap representation of $\dfawitness$, according to Lemma~\ref{lemma:bitmap-witness-sc-dfa}, and~$\aut{A}$ be the minimal \dfa for $\dfawitness$ given by the construction in Algorithm~\ref{algorithm:min-dfa-block}. As we previously saw, the width of rank $i\in[\ell]$ of $\aut{A}$ is given by the cardinality of the set $\bmset{i}$. Therefore, let us analyze the cardinalities of the sets $\{\bmset{i}\}_{i\in[\ell]}$, for the possible values of $\m$.
	
		\begin{enumerate}
		\item $\m=2^{\ell-r}$:
			\begin{enumerate}
				\item\label{item:maxminA} $(\forall j\in [r-1])( |\bmset{j}| = 2^{2^{j}}-1$): \\
					In particular, $\bmset{r}$ contains words of length $2^r$ that correspond to the binary representation of the integers from $1$ to $\m$. For the ranks $j<r$, the set $\bmset{j}$ will contain the binary representation of the same numbers but using $2^j$ bits, as the segments of this set have length $2^j$. Then, we have
				 	$$\bmset{j}=\cset{\reversal{\,\pad(\bin{i},2^j)} \mid \forall i \in [1, \m]}.$$
					Since both $|\bmset{j}| \leq 2^{2^j}-1$ and $\m \geq 2^{2^j}-1$, the proposition holds.
				\item\label{item:maxminB} $(\forall j\in [r, \ell])(|\bmset{j}| = 2^{\ell-j})$: \\
					From the observation in~(\ref{item:maxminA}) referring to the set $\bmset{r}$, it is easy to see that its cardinality is $|\bmset{r}|=2^{\ell-r}$. Consequently, $|\bmset{r+1}| = \frac{|\bmset{r}|}{2} = 2^{\ell-(r+1)}$. It follows that $|\bmset{j}|=2^{\ell-j}$.
			\end{enumerate}
		\item $\m=2^{2^{r-1}}-1$:
			\begin{enumerate}
				\item $(\forall j\in [r-1])( |\bmset{j}| = 2^{2^j}-1)$: \\
					Analogous to the first case~(\ref{item:maxminA}).
				\item $(\forall j\in [r, \ell])( |\bmset{j}| = 2^{\ell-j})$:  \\
				As we saw on the previous case~(\ref{item:maxminB}) it suffices to show that $|\bmset{r}|=2^{\ell-r}$. Recall that, in this case, $\maxr=r-1$. By construction, $\bm(\dfawitness)$ is composed by $\m$ consecutive segments of length $2^{\maxr}$ with at least one bit equal to $1$ followed by a single segment of zeros. Since $\m$ is odd, each element of $\bmset{r}$, consisting of blocks of length $2^r$, is equal either to the binary representation of two consecutive numbers or the second number represented is zero. Therefore, $|\bmset{r}|=2^{2^{\maxr}-1}$ and, as mentioned in proof of Lemma~\ref{lemma:bitmap-witness-sc-dfa}, $2^{r-1}=\ell-r+1$ for this particular values of $\m$. Thus, $|\bmset{r}|=2^{2^{r-1}-1}=2^{\ell-r+1-1}=2^{\ell-r}$, as desired.
			\end{enumerate}
		\end{enumerate}
\end{proof}

\begin{example}
	In Figure~\ref{figure:dfa-witness} the minimal \dfa for $\dfawitnessinst{5}$ is depicted, where $\m=8$ and $\maxr=2$. We have that $\dsc(\dfawitnessinst{5})=2^4-1+2-1+4-1+1=20$.
\end{example}
\begin{figure}[ht]
	\centering
	\begin{tikzpicture}[>=stealth', shorten >=1pt, auto, node distance=1cm,initial text={}, scale=0.8, every node/.style={scale=0.8}]
     	\node[sstate]           (7)                                      {$q_{7}$};
     	\node[sstate]           (8)  [below of=7]                        {$q_{8}$};
     	\node[sstate]           (10) [below of=8]  				         {$q_{10}$};
     	\node[sstate]           (9)  [below of=10]                       {$q_{9}$};
     	\node[sstate]           (11) [below of=9]				         {$q_{11}$};
     	\node[sstate]           (12) [below of=11]  			         {$q_{12}$};
     	\node[sstate]           (13) [below of=12]          	         {$q_{13}$};
     	\node[sstate]           (14) [below of=13]				         {$q_{14}$};
     	\node[sstate]           (3)  [left of=8,xshift=-1cm]	         {$q_{3}$};
     	\node[sstate]           (4)  [below of=3,yshift=-.7cm]	         {$q_{4}$};
     	\node[sstate]           (5)  [below of=4,yshift=-.7cm]	         {$q_{5}$};
     	\node[sstate]           (6)  [below of=5,yshift=-.7cm]	         {$q_{6}$};
     	\node[sstate]           (1)  [left of=4,xshift=-1cm]	         {$q_{1}$};
     	\node[sstate]           (2)  [left of=5,xshift=-1cm]	         {$q_{2}$};
     	\node[sstate,initial]   (0)  [left of=1,xshift=-1cm,yshift=-1cm] {$q_{0}$};
     	\node[sstate]           (15) [right of=10,xshift=2cm]	         {$q_{15}$};
     	\node[sstate]           (17) [right of=11,xshift=2cm]	         {$q_{17}$};
      	\node[sstate]           (16) [right of=13,xshift=2cm]	         {$q_{16}$};
     	\node[sstate,accepting] (18) [right of=17,xshift=1cm]	         {$q_{18}$};
      	\path[->] 
        	(0)  edge                node                       {$a$}   (1)
          	     edge                node[swap]                 {$b$}   (2)
        	(1)  edge                node[pos=.44]              {$a$}   (3)
          	     edge                node[swap,pos=.3]          {$b$}   (4)
        	(2)  edge                node[pos=.36]              {$a$}   (5)
          	     edge                node[swap]                 {$b$}   (6)
        	(3)  edge                node                       {$a$}   (7)
          	     edge                node[swap,pos=.35]         {$b$}   (8)
        	(4)  edge                node[swap,pos=.47]         {$a$}   (9)
          	     edge                node                       {$b$}   (10)
        	(5)  edge                node[pos=.35]              {$a$}   (11)
          	     edge                node[swap]                 {$b$}   (12)
        	(6)  edge                node[inner sep=1,pos=.27]  {$a$}   (13)
          	     edge                node[swap]                 {$b$}   (14)
        	(7)  edge                node[pos=.1]               {$a$}   (15)
       	 	(8)  edge                node[pos=.1]               {$a$}   (16)
        	(9)  edge                node[pos=.1,inner sep=1]   {$a$}   (17)
        	(10) edge                node[pos=.35]              {$b$}   (15)
        	(11) edge                node[pos=.77,inner sep=.2]   {$a,b$} (15)
        	(12) edge                node[pos=.3,inner sep=1]   {$a$}   (15)	
                 edge                node[pos=.1,inner sep=1]   {$a$}   (16)
        	(13) edge[out=30,in=260] node[pos=.1,inner sep=1]   {$b$}   (15)
          		 edge                node[pos=.38,inner sep=1]  {$a$}   (16)
        	(14) edge                node[swap] [pos=.3]        {$b$}   (16)
        	(15) edge                node[inner sep=1,pos=.3]   {$a$}   (18)
        	(16) edge                node[inner sep=1,pos=0.45] {$b$}   (18)
        	(17) edge                node[inner sep=1,pos=.3]   {$a,b$} (18)  
        	;
		\path[gray]
			(14)++(-90:5ex) node (rank 2) {rank $2$}
			(rank 2 -| 6)   node (rank 3) {rank $3$}
			(rank 2 -| 2)   node (rank 4) {rank $4$}
			(rank 2 -| 0)   node (rank 5) {rank $5$}
			(rank 2 -| 16)  node (rank 1) {rank $1$}
			(rank 2 -| 18)  node (rank 0) {rank $0$}
		;
		\foreach \r [remember=\r as \lastr (initially 0)] in {1,...,5} {
			\draw[dashed,gray] ($(rank \r)!.5!(rank \lastr)$)+(0,8.2cm) -- ($(rank \r.south)!.5!(rank \lastr.south)$);
		}
	\end{tikzpicture}
	\caption{The minimal \dfa accepting the language $\dfawitnessinst{\ell}$ for $\ell=5$.}
	\label{figure:dfa-witness}
\end{figure}

Using the result from Câmpeanu and Ho~\cite{campeanu04} on the number of states of a minimal DFA accepting a block language $L\subseteq\Sigma^\ell$, along with the approximation of one of its parameters in Theorem~\ref{theorem:sc-dfa}, we obtain the following asymptotic approximation for~$\dsc(L)$:
\begin{corollary}
	Given a block language $L\subseteq \alphabet^\ell$, with $\ell >0$ and $|\alphabet|=k$, the asymptotic behaviour of $\dsc(L)$ is $\Theta(\frac{k^\ell}{\ell} + 2^\ell)$.
\end{corollary}

\subsection{Maximal Size of Minimal NFAs} \label{sec:sc-nfa}
Let $L\subseteq\alphabet^\ell$ be a block language of block length $\ell >0$ and $k=|\alphabet|$, with bitmap representation $\bm$. Let $\aut{A}$ be a minimal \nfa for $L$, given by the construction given in Section~\ref{sec:nfa-construction}. The states in rank $i$ of $\aut{A}$ are given by the set $\cover{i}$, a minimal cover for the set of segments of length $k^i$ of $\bm$, namely $\bmset{i}$, for $i\in[\ell]$. The size of the set $\cover{i}$ is bounded by the size of $\bmset{i}$. Also, it can be noticed that $k^i$ words suffice to cover any set $\bmset{i}$. These bounds are formally stated in the following lemma:

\begin{lemma} \label{lemma:size-of-covers}
	Let $L\subseteq\alphabet^\ell$ be a block language of words of length $\ell >0$ over a $k$-letter alphabet $\alphabet$, with a correspondent bitmap $\bm$. Then, for each $i\in[\ell]$, the cardinality of~$\cover{i}$, a minimal cover for $\bmset{i}$, is bounded by $|\cover{i}| = min(k^{\ell-i}, k^i)$.
\end{lemma}

\begin{proof}
	The bound $|\cover{i}|\leq k^{\ell-i}$, for $i\in[\ell]$, refers to the maximal size of $\bmset{i}$ (Lemma~\ref{lemma:size-of-segments}). As we previously saw, $\bmset{i}$ covers itself so the bound follows. On the other hand, we have $|\cover{i}|\leq k^i$ since the set $\bmset{i}$ can be covered by the set of unit segments $\{u_j\}_{j\in[k^i-1]}$, where $u_j$ represents the word of zeros, apart from the $j$-th bit, which is $1$, and $|u_j|=k^i$.
\end{proof}

We say that a minimal \nfa is of maximal size if, for each $i\in[\ell]$, the width of rank $i$ is $min(k^{\ell-i}, k^i)$. As a result of the previous lemma, we are able to determine an upper bound for the nondeterministic state complexity of a block language.

\begin{theorem}\label{theorem:max-nfa}
The maximal size of a minimal \nfa for a block language $L\subseteq \alphabet^\ell$, with $\ell >0$ and $|\alphabet|=k$, is
\begin{align*}
		\nsc(L) \leq 
		\begin{cases}
			{\displaystyle 2\cdot\frac{k^{\frac{\ell}{2}}-1}{k-1} + k^{\frac{\ell}{2}}}, & \text{if } \ell \text{ is even;} \\[20pt]
			{\displaystyle 2\cdot \frac{k^{\lceil \frac{\ell}{2}\rceil} -1}{k-1}}, & \text{otherwise.}
		 \end{cases}
	\end{align*}
\end{theorem}
\begin{proof}
	If $\ell$ is even, then there is an odd number of ranks, and the width of the minimal NFA with the maximal size is attained at the rank $j\in[\ell]$ that satisfies $k^{\ell-j}=k^j$, according to Lemma~\ref{lemma:size-of-covers}. This implies $j=\frac{\ell}{2}$, so we have:
		$$ \nsc(L) \leq 2\cdot\sum_{i=0}^{\frac{\ell}{2}-1} k^i + k^{\frac{\ell}{2}} = 2\cdot\frac{k^{\frac{\ell}{2}}-1}{k-1} + k^{\frac{\ell}{2}}.$$	
	If $\ell$ is odd, there are an even number of ranks, and the width of the minimal NFA with maximal size is reached both in rank $\lfloor\frac{\ell}{2}\rfloor$ and $\lceil\frac{\ell}{2}\rceil$. Then, we have:
	$$ \nsc(L) \leq 2\cdot\sum_{i=0}^{\lfloor\frac{\ell}{2}\rfloor} k^i = 2\cdot\frac{k^{\lfloor\frac{\ell}{2}\rfloor+1}-1}{k-1} = 2\cdot\frac{k^{\lceil\frac{\ell}{2}\rceil}-1}{k-1},$$
	as desired.
\end{proof}

We shall now present a family of \emph{witness languages} which minimal \nfas that recognize them are of maximal size. Let $k\geq2$ and $d\geq 0$. Then, the languages
	$$ L_{k,d} = \{\,w\reversal{w} \mid w\in\alphabet^d\,\},$$
	defined over $\alphabet=\{\letter_0,\ldots,\letter_{k-1}\}$, correspond to the set of palindromes of length~$2d$.
	
\begin{lemma} \label{lemma:nsc-nfa}
	Every minimal NFA that recognizes the language $L_{k,d}$ is of maximal size,
	where $k>1$ and $d\geq 0$.
\end{lemma}
\begin{proof}
Let $\aut{A}$ be a minimal \nfa for $L_{k,d}$, and $m_i$ be the width of $\aut{A}$ at rank $i\in[2d]$. Let us prove that $m_i$ is maximal, that is:
	\begin{enumerate}
		\item $(\forall i\in[d,2d])( m_i=k^{2d-i}):$ \label{item:nsc-nfa-1} \\
		Let $w_1,w_2\in\alphabet^{2d-i}$ be two words that differ by at least one symbol, for $i\in [d,2d]$. Let us prove that $\derivative{w_1}L_{k,d}\cap\derivative{w_2}L_{k,d}=\emptyset$. Let $w_3\in\alphabet^i$ such that $w_1w_3\in L_{k,d}$. Then, $w_3$ can be decomposed into $u\reversal{u}\reversal{w}$, such that $|u|=\alphabet^{i-d}$. On the other hand, $w_2w_3\in L_{k,d}$ if and only if $w_1=w_2$, which is a contradiction. Thus, $\derivative{w_1}L_{k,d}\cap\derivative{w_2}L_{k,d}=\emptyset$ and, as a consequence, $k^{2d-i}$ states are needed at rank~$i$, one for each word in $\alphabet^{2d-i}$.
		\item $(\forall i\in[d])( m_i=k^i):$ \\
		Let us look at $\reversal{\aut{A}}$, the NFA for $\reversal{L_{k,d}}$ given by reversing every transition in $\aut{A}$ and swapping the initial with the final state. In fact, it is easy to see that $L_{k,d}=\reversal{L_{k,d}}$, hence $\lang{A}=\lang{\reversal{\aut{A}}}$. In (\ref{item:nsc-nfa-1}), we proved that $m_j=k^{2d-j}$, for every rank $j\in[d,2d]$. The $i$-th rank in $\aut{A}$ corresponds to the $(2d-i)$-th rank in $\reversal{\aut{A}}$, so that bound must be preserved.
	\end{enumerate}
\end{proof}

By the arguments provided in the previous proof, we have the following corollary.
\begin{corollary} \label{corollary:nsc-nfa}
	The minimal NFA for $L_{k,d }$ is deterministic.
\end{corollary}

From Theorem~\ref{theorem:max-nfa}, we obtain the following asymptotic approximation for $\nsc(L)$, where $L$ is a block language:
\begin{corollary}
	Given a block language $L\subseteq \alphabet^\ell$, with $\ell >0$ and $|\alphabet|=k$, the asymptotic behaviour of $\nsc(L)$ is $\Theta(\sqrt{k^\ell})$.
\end{corollary}

\subsection{Operational State Complexity} \label{sec:osc}
In this section we consider operations on block languages using their bitmap representations and study both the deterministic and nondeterministic state complexity of these operations. More precisely, the \emph{operational state complexity} is the worst-case state complexity of a language resulting from the operation, considered as a function of the state complexities of the operands. For instance, the state complexity of the union of two block languages can be stated as follows: given an $m$-state \dfa~$\aut{A}_1$ and an $n$-state \dfa~$\aut{A}_2$, how many states are sufficient and necessary, in the worst case, to accept the language $L(\aut{A}_1)\cup L(\aut{A}_2)$ by a \dfa? 

An upper bound can be obtained by providing an algorithm that, given the \dfas for the operands, constructs a \dfa that accepts the resulting language. The number of states, in the worst case, of the resulting \dfa is an upper bound for the state complexity of the referred operation. To show that an upper bound is tight, a family of languages (one language, for each possible value of the state complexity) for each operation must be given such that the resulting automata achieve that bound. We call those families \emph{witnesses} or \emph{streams}.
 
Of course, the upper bounds of operational state complexity known for finite languages apply for block languages. In Table~\ref{tab:cfin}, we review some complexity results for finite languages~\cite{gao17}. The first two lines give the bounds for the determinization of an $m$-state \nfa and the asymptotic upper bound of the maximal size of a minimal \dfa, respectively. When considering unary operations, let $L$ be a finite language with~$\dsc(L)=m$ ($\nsc(L)=m$) and let $\aut{A}=\auttuple{Q}{\alphabet}{\delta}{q_0}{F}$ be the complete minimal \dfa (a minimal NFA) for $L$. Furthermore, we consider $k=|\alphabet|$ or $|\alphabet|=f(\overline{m})$ if a growing alphabet is taken into account, $f=|F|$, and $p=|F-\{q_0\}|$. In the same way, for binary operations let $L_1$ and $L_2$ be finite languages over the same alphabet with $\dsc(L_1)=m$ ($\nsc(L_1)=m$) and $\dsc(L_2)=n$ ($\nsc(L_2)=n$), and let $\aut{A}_i=\auttuple{Q_i}{\alphabet}{\delta_i}{q_i}{F_i}$ be complete minimal DFAs (minimal NFAs) for $L_i$, with~$i\in\{1,2\}$. Also, $p_i=|F_i-\{q_i\}|$.

\begin{table}
	\centering
{	\begin{tabular}{lcccccc}
		\toprule
		& \multicolumn{1}{c}{Upper bound} & \multicolumn{1}{c}{$|\alphabet|$} \\
		\midrule
		\nfa $\to$ \dfa & $\Theta(k^{\frac{m}{1+\log k}})$ & $2$ \\
		$\dsc(L)$ & $\frac{k^{\ell+2}}{\ell(k-1)^2\log_2{k}}(1+o(1))$ & $2$ \\
		\midrule
		& \multicolumn{1}{c}{$\dsc$} & \multicolumn{1}{c}{$|\alphabet|$} & \multicolumn{1}{c}{$\nsc$} & \multicolumn{1}{c}{$|\alphabet|$} \\
		\midrule
		$L_1\cup L_2$ & $mn-(m+n)$ & $f(m,n)$ & $m+n-2$ & $2$ \\
		$L_1\cap L_2$ & $mn - 3(m + n) + 12$ & $f(m,n)$ & $O(mn)$ & $2$ \\
		$\overline{L}$ & $m$ & $1$  & $\Theta(k^{\frac{m}{1+\log k}})$ & $2$ \\ \addlinespace[2mm]
		\multirow{2}{*}{$L_1L_2$} &$(m-n+3)2^{n-2}-1$, $m+1\geq n$ & $2$ & \multirow{2}{*}{$m+n-1$} & \multirow{2}{*}{$2$} \\
		& $m+n-2$, if $p_1=1$ & $1$ & & \\ \addlinespace[2mm]
		\multirow{2}{*}{$\star{L}$} & $2^{m-3}+2^{m-p-2}$, $p\geq 2$, $m\geq 4$ & $3$ & \multirow{2}{*}{$m-1$, $m>1$} & \multirow{2}{*}{$1$} \\ & $m-1$, if $f=1$ & $1$ & & \\ \addlinespace[2mm]
		$L^{R}$ & $O(k^{\frac{m}{1+\log k}})$ & $2$ & $m$ & $2$ \\
		\bottomrule
	\end{tabular}}
	\caption{Some complexity bounds for finite languages} 
	\label{tab:cfin}
\end{table}

We mainly consider operations under which the family of block languages is closed, i.e., the resulting language is also a block language. These include the reversal, the word addition and removal from a block language, the intersection, the union, the concatenation, and the complementation of block languages closed to the block (i.e., $\alphabet^\ell \setminus L$). We shall also analyze the Kleene star and plus on block languages, as well as the block language stencil operation, which in general do not yield a block language.

Additionally, a part from the Boolean operations already considered, we show how to build the bitmap of the language resulting by applying each operation. We  also present a family of witness languages parameterized by the state complexity of the operands to show that the bounds provided are tight. In general, other additional parameters are the  block length $\ell$ of the words and the widths of the ranks.

\subsubsection{Reversal} \label{sec:reversal}
In the following, given a bitmap $\bm$ of a block language $L\subseteq\alphabet^\ell$, $|\alphabet|=k$, and $\ell>0$, we compute the bitmap for the reversal language $\reversal{L}$, namely $\reversal{\bm}$. The bitmap for $\reversal{L}$ can be constructed by swapping the values of $b_i$ and $b_j$ from $\bm$, such that $i=\ind(w)$ and $j=\ind(\reversal{w})$, for some $w\in\alphabet^\ell$. Trivial solutions require $k^\ell$ additional space to either store in a set the indexes already swapped or to construct a new bitmap. We now propose a routine that constructs the bitmap $\reversal{\bm}$ in-place, that is, by performing operations on $\bm$ and without requiring extra space. A similar technique is also described in~\cite{Diaconis:1983aa}.

Recall that the perfect shuffle of length $1$ of two words $u=u_0\cdots u_{n-1}$ and $v=v_0\cdots v_{n-1}$ of the same length $n$, denoted by $u\shuffle_1v$, is obtained by interleaving the letters of $u$ and $v$, namely, 
$$u\shuffle_1 v= u_0v_0\cdots u_{n-1}v_{n-1}\text.$$ 
If $j\in\naturals$ is a divisor of $n$, the perfect shuffle of length $j$, denoted as $\shuffle_j$, of $u$ and $v$ is the perfect shuffle of blocks of length $j$, that is, 
$$u \shuffle_j v = u_0\cdots u_{j-1}v_0\cdots v_{j-1} \cdots u_{n-j}\cdots u_{n-1} v_{n-j}\cdots v_{n-1}\text.$$
Moreover, this operation can be extended for every number $m\geq 2$ of words $w_0,\ldots, w_{m-1}$ of the same length $n$. We define $\shuffle^m_j(w_0\cdots w_{m-1})$ as the perfect shuffle of blocks of length $j$ taken from each of the $w_i$ words, where $j$ is a divisor of $n$. That is, with $w_i=w_{i,0}\cdots w_{i,n-1}$, for each $i\in[m-1]$,
	$$ \shuffle_j^m(w_0\cdots w_{m-1}) = \prod_{r=0}^{\frac{n}{j}-1}\left(\prod_{i=0}^{m-1}w_{i,rj}\cdots w_{i,(r+1)j}\right)\text.$$
For instance, for $j=1$ we have,
$$ \shuffle_1^m(w_0\cdots w_{m-1})=w_{0,0}\cdots w_{m-1,0}w_{0,1}\cdots w_{m-1,1}\cdots w_{0,n-1}\cdots w_{m-1,n-1}.$$

Let us now show that the shuffle operation can be used to obtain the bitmap representation of the reversal of a language. Given a bitmap $\bm$ of length $k^\ell$, with~$k>1$ and $\ell>0$, we define $\rotation_i$, for each $i\in[\ell-1]$, as follows:
\begin{align*}
	\rotation_i =
	\begin{cases}
		\bm,                                    & \text{if } i=0; \\
		\shuffle^k_{k^{i-1}}(\mathsf{R}_{i-1}), & \text{otherwise.}
	\end{cases}
\end{align*}

\begin{lemma}
	Let $L\subseteq\alphabet^\ell$ be a block language, for some $\ell>0$ and $|\alphabet|=k$, with bitmap representation $\bm$. The bitmap for the reversal of $L$, namely $\reversal{\bm}$, is $\mathsf{R}_{\ell-1}$.
\end{lemma}
\begin{proof}
	Let us prove that $\lang{\mathsf{R}_{\ell-1}}=\reversal{L}$. For $i=0$, we have~$\lang{\mathsf{R}_0}=\lang{\bm}=L$. Next, for $i=1$, we have $\mathsf{R}_1=\shuffle^k_1(\bm)$. This operation performs the cyclic permutation $S_1=(0\;1\;\cdots\;\ell-1)$ in each word of $\lang{\bm}$, that is, each symbol of every word in $\lang{\bm}$ is shifted one position to their right and the last symbol becomes the first. The following operation, $\mathsf{R}_2=\shuffle^k_k(\mathsf{R}_1)$, performs the permutation $S_2=(1\;2\;\cdots\;\ell-1)$ in every word of $\lang{\mathsf{R}_1}$. Analogously, in this transformation each symbol apart from the first of every word in $\lang{\mathsf{R}_1}$ is shifted one position to their right but the last symbol now becomes the second. In general, the $j$-th shuffle performs the permutation $S_j=(j-1\;j\;\cdots\;\ell-1)$, for $j\in [1,\ell-1]$. The composition of the transformations $S_1,S_2,\ldots,S_{\ell-1}$ ensure that $\lang{\mathsf{R}_{\ell-1}}=\reversal{L}$~\cite{Diaconis:1983aa}.
\end{proof}

\begin{example}
	Let $\alphabet=\{a,b\}$ and $\ell=3$. Let $\bm=b_1b_2b_3b_4b_5b_6b_7b_8$ be a bitmap for a block language $L$ such that $b_1=b_4=b_5=1$, and the remaining bits are $0$. We have 
	\begin{align*}
		\rotation_0 & = b_1b_2b_3b_4b_5b_6b_7b_8 \quad \text{ and } \quad \lang{\rotation_0} = \{aaa, abb, baa\}, \\
		\rotation_1 & = b_1b_5b_2b_6b_3b_7b_4b_8 \quad \text{ and } \quad \lang{\rotation_1} = \{aaa, bab, aba\}, \\
		\rotation_2 & = b_1b_5b_3b_7b_2b_6b_4b_8 \quad \text{ and } \quad \lang{\rotation_2} = \{aaa, bba, aab\},
	\end{align*}
	and $\lang{\rotation_2}=\reversal{L}$, as desired.
\end{example}

Now we turn to the analysis of the state complexity of this operation. 
\begin{lemma} \label{lemma:sc-reversal-sufficiency}
	Given an $m$-state \dfa for a block language $L$, $2^{O(\sqrt{m})}$ states are sufficient for a \dfa accepting $\reversal{L}$.
\end{lemma}
\begin{proof}
	A \dfa for $\reversal{L}$, with $\ell>0$, can be given by reversing each transition of the \dfa for $L$, swapping the initial and final states, and then determinizing the resulting NFA. The cost of the determinization of an $m$-state NFA for a block language is~$2^{\Theta(\sqrt{m})}$ in terms of number of states~\cite{KarhOkho:2O14}, so the state complexity of the reversal must also be limited by this bound. 
\end{proof}

In the following, we show that this bound is tight. Recall the language $\dfawitness$, defined in Section~\ref{sec:sc-dfa}, over the alphabet $\alphabet=\{a,b\}$ and for $\ell>0$. As before, let
\begin{enumerate}
	\item $r=\min\{\,i\in[\ell]\mid 2^{\ell-i}\leq 2^{2^{i}}-1\,\}$, where $r = \lfloor \log_k\ell \rfloor + 1 + x$ for some  $x\in\{-1,0,1\}$;
	\item $\m=\max(2^{\ell-r}, 2^{2^{r-1}}-1)$;
	\item $\maxr=r$, if $\m=2^{\ell-r}$, or $\maxr=r-1$, if $\m=2^{2^{r-1}}-1$.
\end{enumerate}

We claim that the minimal \dfa for the reversal of $\dfawitness$, namely $\reversal{\dfawitness}$, has its width bounded by $2^{r_\ell+1}$. That is, for every rank $i\in[\ell]$, $\width(i)\leq 2^{r_\ell+1}$.
\begin{lemma} \label{lemma:sc-reversal-maxrank}
	A minimal \dfa~$\aut{A}$ such that $\width(\aut{A})\leq 2^{r_\ell+1}$ is sufficient to accept the reversal of the language $\dfawitness$, for $\ell>0$.
\end{lemma}
\begin{proof}	
	Let $\aut{A}$ be a \dfa such that the last $\maxr+1$ ranks have maximal width (according to Theorem~\ref{theorem:sc-dfa}), that is, $\width(i)=2^{\ell-i}$, for each $i\in [\ell-\maxr, \ell]$. In particular, we have that the width of the rank $\ell-\maxr$ is $2^{\maxr}$. We can order the states in this rank in such a way that $q_j$ is the state whose left language is the reverse of the $j$-th word of $\alphabet^{r_\ell}$, for each $j\in[2^{\maxr}-1]$. Formally, $\stateleftlang{\aut{A}}{q_j}=\{\reversal{w}\}$ such that $|w|=\maxr$ and $\ind(w)=j$. Moreover, let us define the right language of $q_j$ as
		$$\staterightlang{\aut{A}}{q_j} = \{\,w\in\alphabet^{\ell-r_\ell} \mid i=\ind(\reversal{w}), \bit(\bin{(i-1)},j)=1\,\}.$$ 		
		From the definition of the language, it is easy to see that $\aut{A}$ accepts $\reversal{\dfawitness}$. Consider~$\aut{A}_{q_j}$ as the DFA $\aut{A}$ with initial state $q_j$, and let us show that $\width(\aut{A}_{q_j}) \leq 2$, for all $j\in[2^{\maxr}-1]$.

	Let $j\in[2^{\maxr}-1]$, $r'\in[\ell-\maxr-1]$, and $\bmset{r'}$ be the set of segments of size $2^{r'}$ of the bitmap of $\reversal{\dfawitness}$. Let $s\in\bmset{r'}$ and $w_1\in\alphabet^{\ell-\maxr-r'}$ such that $s$ represents the language $\derivative{w_1}\staterightlang{\aut{A}}{q_j}$, according to Lemma~\ref{lemma:segments}. From the definition of $\staterightlang{\aut{A}}{q_j}$, $w_1w_2\in\staterightlang{\aut{A}}{q_j}$ if and only if $\bit(\bin{(i+1)},j)=1$, where $w_2\in\alphabet^{r'}$ and $i=\ind(\reversal{(w_1w_2)})=\ind(\reversal{w_2}\reversal{w_1})$. We shall now show that the number of states on the rank $r'$ of $\aut{A}_{q_j}$ is bounded by $2$, by arguing that $|\bmset{r'}| \leq 2$. Consider the following two cases:
		\begin{enumerate}
			\item $j < \ell-r_\ell-r'$: \\
				In this case, the membership of $w_1w_2$ in $\staterightlang{\aut{A}}{q_j}$ has already been decided by the choice of $w_1$. Then, it is sufficient to check whether~$\bit(\bin{(i'+1)},j)=1$, where $i'=\ind(\reversal{w_1})$, since $|\reversal{w_1}|=\ell-r_\ell-r'$. Then, either $\bin{(i'+1)}$ has its $j$-th bit ($j$-th but last symbol) equal to $1$, which implies that $s=1\cdots 1$, or has not, implying that $s=0\cdots 0$, so $s\notin\bmset{r'}$. Therefore, $|\bmset{r'}| = 1$.
			\item $j \ge \ell-r_\ell-r'$: \\			
				Now, $w_1w_2\in\staterightlang{\aut{A}}{q_j}$ depends on the choices of $w_1$. If $w_1\in\alphabet^{\ell-\maxr-r'}\setminus\{b^{\ell-\maxr-r'}\}$, the binary representation of $i'+1$, with $i'=\ind(\reversal{w_1})$, requires at most $\ell-r_\ell-r'$ bits. Then, the $j$-th bit ($j$-th but last symbol) of $i$, corresponds to the ($j-\ell+\maxr+r'$)-th bit of $w_2$. Hence, $\derivative{u}\staterightlang{\aut{A}}{q_j} = \derivative{v}\staterightlang{\aut{A}}{q_j}$, for all $u,v\in\alphabet^{\ell-\maxr-r'}\setminus \{b^{\ell-r_\ell-r'}\}$. On the other hand, if $w_1=b^{\ell-\maxr-r'}$, it results on a different quotient, since $\ell-\maxr-r'+1$ bits are needed for the binary representation of $\ind(w_1)+1$. Therefore, $|\bmset{r'}| = 2$.
		\end{enumerate}
	This result implies that $\width(\aut{A}_{q_j})=2$. As a consequence, the width of the ranks $r'\in [\ell-r_\ell-1]$ of $\aut{A}$ are bounded by $2^{\maxr+1}$, as desired.
\end{proof}

We now show that every~\dfa for $\dfawitness$ is exponentially larger than the minimum \dfa for $\reversal{\dfawitness}$.
\begin{lemma} \label{lemma:sc-reversal-necessary}
	Let $\aut{A}_1$ be an $m$-state \dfa for $\reversal{\dfawitness}$. Then, every DFA $\aut{A}_2$ for $\dfawitness$ needs at least $2^{\Omega(\sqrt{m})}$ states.
\end{lemma}
\begin{proof}
	For this proof, assume that $2^{\ell-r} > 2^{2^{r-1}}-1$ (i.e., $\m=2^{\ell-r}$), which implies that $\maxr=r$. A similar proof follows, otherwise.

	Let $Q$ and $P$ be the sets of states of $\aut{A}_1$ and $\aut{A}_2$, respectively. By Lemma~\ref{lemma:witness-maxminDFA}, $\aut{A}_2$ is of maximal size and $\width(\aut{A}_2)=2^{\ell-r}=\m$. Therefore, the number of states of $\aut{A}_2$ is
	\begin{align*}
		|P| & = \sum_{i=0}^{r_\ell-1}(2^{2^i}-1) + \sum_{i=r_\ell}^{\ell}2^{\ell-i} \\
		    & = \sum_{i=0}^{r_\ell-1}2^{2^i} - r_\ell + 2^{\ell-r_\ell+1} - 1 \\
		    & \geq 2^{2^{r_\ell}-1} - (r_\ell+1) \tag*{$2^{\ell-r_\ell+1} > 0$} \\
		    & \geq 2^{2^{\log_2\ell+2}-1} - (\log_2\ell+3) \tag*{$2^{2^{r_\ell}-1} \gg r_\ell$, $r=r_\ell$, and $r<\log_2\ell+2$} \\
			& \geq 2^{4\ell-1} - (\log_2\ell+3) \in 2^{\Omega(\ell)}. \\
	\end{align*}
	By Lemma~\ref{lemma:sc-reversal-maxrank}, the size of $\aut{A}_1$ is bounded by $2^{\maxr+1}$. Moreover, the width of each rank $i\in[r-1]$ of $\aut{A}_1$ is also bounded by $2^{2^i}-1$, as we have seen in Theorem~\ref{theorem:sc-dfa}. Then, let $r'=\min\{\,i\in[r-1] \mid 2^{r_\ell+1} \leq 2^{2^i}-1\,\}$. In particular, we have
	\begin{align*}
		2^{r_\ell+1} > 2^{2^{r'-1}}-1 & \implies 2^{r_\ell+1} \geq 2^{2^{r'-1}} \\
									  & \implies r_\ell+1 \geq 2^{r'-1} \\
									  & \implies \log_2\ell + 3 \geq 2^{r'-1} \tag*{$r=r_\ell$ and $r<\log_2\ell+2$} \\
									  & \implies r' \leq \log_2(\log_2\ell +3) + 1.
	\end{align*}
	The value of $r'$ tells us how many ranks in $\aut{A}_1$ can achieve the maximal width of $2^{r_\ell+1}$. Then, the number of states of $\aut{A}_1$ is bounded by
	\begin{align*}
		|Q| & \leq \sum_{i=0}^{r'-1}(2^{2^i}-1) + 2^{r_\ell+1}(\ell-r_\ell-r') + \sum_{i=\ell-r_\ell}^{\ell}2^{\ell-i} \\
			& \leq 2^{2^{r'}} - r' + 2^{r_\ell+1}(\ell-r_\ell-r'+1) - 1 \\ 
			& \leq 2^{2^{\log_2(\log_2\ell +3) + 1}} + 2\cdot2^{r_\ell}(\ell+1) \tag*{$r' \leq \log_2(\log_2\ell +3) + 1$} \\
		    & \leq 2^6\cdot2^{\log_2\ell^2} + 2^3\cdot2^{\log_2\ell}(\ell+1) \tag*{$r=r_\ell$ and $r<\log_2\ell+2$} \\
			& \leq 2^6\ell^2 + 2^3(\ell^2+\ell) = O(\ell^2). \\
	\end{align*}
	Thus, we have that $\dsc(\reversal{\dfawitness})=m=O(\ell^2)$ and $\dsc(\dfawitness)\in2^{\Omega\left(\sqrt{m}\right)}$, as desired.
\end{proof}

With the results in Lemmas~\ref{lemma:sc-reversal-sufficiency} and~\ref{lemma:sc-reversal-necessary}, we have:
\begin{theorem}\label{theorem:sc-reversal}
	Let $L\subseteq\alphabet^\ell$, for some $\ell>0$, such that $\dsc(L)=m$. Then, $\dsc(\reversal{L})\in2^{O(\sqrt{m})}$, and the bound is tight.
\end{theorem}

The \nfa for the reversal of a language $L\subseteq\alphabet^\ell$ is given by reversing the transitions on the \nfa  for $L$ and swapping the initial and final states. In fact, the nondeterministic state complexity of the reversal of a finite language coincides with the nondeterministic state complexity of the language, so no better result can be obtained for block languages.
\begin{theorem}\label{theorem:nsc-reversal}
	Let $L\subseteq\alphabet^\ell$, for some $\ell>0$. Then, $\nsc(\reversal{L})=\nsc(L)$.
\end{theorem}
\begin{proof}
	The construction above shows that $\nsc(\reversal{L})\leq\nsc(L)$. The following family of languages shows that it is tight. Let $L_\ell=\{a^\ell\}$, with $\ell>0$. Since $L_\ell=\reversal{L_\ell}$, $\nsc(L_\ell)=\nsc(\reversal{L_\ell})$.
\end{proof}

\subsubsection{Word Addition and Word Removal} \label{sec:word-op}
Consider a language $L\subseteq\alphabet^\ell$, for some $\ell>0$, over an alphabet of size $k$. The operations of adding or removing a word $w\in\alphabet^\ell$ from the language, $L\setminus\{w\}$ and $L\cup\{w\}$, respectively, corresponds to the not operation on the $\ind(w)$-th bit of $\bm$, the bitmap of $L$, provided that the resulting language is different from $L$. From that observation, we can estimate the state complexity of these operations.

\begin{theorem} \label{theorem:sc-word-op}
	Let $L\subseteq\alphabet^\ell$ be a block language such that $\dsc(L) = m$. Let $L'=L\oplus\{w\}$, for $\oplus \in \{\setminus, \cup\}$ and $w\in\alphabet^\ell$. Then, $m-(\ell-1)\leq\dsc(L')\leq m+(\ell-1)$. 
\end{theorem}
\begin{proof}
	Let $\bm$ and $\bm'$ be the bitmap representations of $L$ and $L'$, respectively. Let us assume that the operation $\oplus$ results in a different language. Then, the bitmaps $\bm$ and $\bm'$ differ exactly by one bit. Moreover, for every $i\in[\ell]$, there is exactly one $j\in[1,k^{\ell-i}]$ such that $\bmsegment{i}{j}\neq t^i_j$, where $\bmsegment{i}{j}$ and $t^i_j$ denote the $j$-th bitmap segment of size~$k^i$ of $\bm$ and $\bm'$, respectively. Also, recall $\bmset{i}$ (resp. $\bmset{i}'$), the set of segments of size~$k^i$ of $\bm$ (resp. $\bm'$). Then, there are four possible cases:
	\begin{enumerate}
		\item $\bmsegment{i}{j}\in\bmset{i}'$ and $t^i_j\in\bmset{i}$: the two sets have the same size;
		\item $\bmsegment{i}{j}\in\bmset{i}'$ and $t^i_j\notin\bmset{i}$: $\bmset{i}'$ has one more element than $\bmset{i}$; \label{item:sc-word-op-increase}
		\item $\bmsegment{i}{j}\notin\bmset{i}'$ and $t^i_j\in\bmset{i}$: $\bmset{i}$ has one more element than $\bmset{i}'$; \label{item:sc-word-op-decrease}
		\item $\bmsegment{i}{j}\notin\bmset{i}'$ and $t^i_j\notin\bmset{i}$: the two sets have the same size.
	\end{enumerate}
	Therefore, the difference on the number of states from a minimal \dfa that accepts the language $L'$ and the \dfa that accepts $L$ is bounded by $\ell-1$, which is the number of ranks neither initial nor final.
\end{proof}

These bounds also extend to the nondeterministic state complexity, as proved in the following result.
\begin{theorem} \label{theorem:nsc-word-op}
	Let $L\subseteq \alphabet^\ell$ be a block language such that $\nsc(L) = m$. Let $L'=L\oplus\{w\}$, for  $\oplus\in\{\setminus, \cup\}$ and $w \in \alphabet^\ell$. Then, $m-(\ell-1)\leq\nsc(L')\leq m+(\ell-1)$.
\end{theorem}
\begin{proof}
	Consider the proof of Theorem~\ref{theorem:sc-word-op} and its notation. If the case (\ref{item:sc-word-op-increase}) verifies, that is, $|\bmset{i}'|=|\bmset{i}|+1$, then the cover requires, at most, one more segment to cover the new set. Analogously, the size of the cover for $\bmset{i}'$ can be smaller by one than the cover for $\bmset{i}$, for the case (\ref{item:sc-word-op-decrease}).
\end{proof}

\begin{theorem}\label{theorem:sc-word-op-witness}
	The bounds given in Theorems~\ref{theorem:sc-word-op} and~\ref{theorem:nsc-word-op} are tight.
\end{theorem}
\begin{proof}
	Let $\alphabet = \{a,b\}$ and $\ell>0$. For word removal, consider $L_\ell=(a+b)^\ell$ and let~$w=a^\ell$. We have $\dsc(L_\ell)=\nsc(L_\ell)+1=\ell+2$.
	One can see that
$$\dsc(L_\ell\setminus\{w\})=\nsc(L_\ell\setminus\{w\})+1=2\ell+1,$$ 
as $\ell-1$ more states are needed in the minimal automata when reading an $a$ in the first $\ell$ letters.	
	Figure~\ref{figure:sc-word-op-witness} shows the minimal \dfa for $\ell=4$. For word addition, consider $L_\ell'=\{a^\ell\}$ and let $w=b^\ell$. We have $\dsc(L_\ell')=\nsc(L_\ell')+1=\ell+2$, while $\dsc(L_\ell'\cup\{w\})=\nsc(L_\ell'\cup\{w\})+1=2\ell+1$.
\end{proof}

\begin{figure}[ht]
	\centering
	\def\shift{1cm}
	\begin{tikzpicture}[>=stealth', shorten >=1pt, auto, node distance=2cm, initial text={},every node/.style={scale=0.8}]
		\node[sstate,initial]                                       (0)    {$q_0$};
		\node[sstate,above of=0,right of=0,yshift=-\shift]          (1)    {$q_1$};
		\node[sstate,below of=0,right of=0,yshift=\shift]           (2)    {$q_2$};
		\node[sstate,right of=1]                                    (3)    {$q_3$};
		\node[sstate,right of=2]                                    (4)    {$q_4$};
		\node[sstate,right of=3]                                    (5)    {$q_5$};
		\node[sstate,right of=4]                                    (6)    {$q_6$};
		\node[sstate,accepting,below of=5,right of=5,yshift=\shift] (7)    {$q_7$};
		\node[sstate,right of=7,xshift=.5\shift]                    (sink) {$\Omega$};

		\path[->]
			(0) edge                node                             {$a$}   (1)
				edge                node[swap]                       {$b$}   (2)
			(2) edge                node[swap]                       {$a,b$} (4)
			(4) edge                node[swap]                       {$a,b$} (6)
			(6) edge                node[swap,inner sep=.5, pos=.25] {$a,b$} (7)
			(1) edge                node[pos=.3]                     {$a$}   (3)
				edge                node[inner sep=.5,pos=.25]       {$b$}   (4)
			(3) edge                node[pos=.3]                     {$a$}   (5)
				edge                node[inner sep=.5,pos=.25]       {$b$}   (6)
			(5)	edge                node[inner sep=.5,pos=.52]       {$b$}   (7)
				edge[bend left=15]  node                             {$a$}   (sink) 
			(7) edge                node[inner sep=.5, pos=.45]      {$a,b$} (sink)
			(sink) edge[loop above] node                             {$a,b$} (sink)
			;
			
		\path[gray]
				(4)++(-90:5ex) node (rank 2) {rank $2$}
				(rank 2 -| 2)  node (rank 3) {rank $3$}
				(rank 2 -| 0)  node (rank 4) {rank $4$}
				(rank 2 -| 6)  node (rank 1) {rank $1$}
				(rank 2 -| 7)  node (rank 0) {rank $0$}
		;
		\foreach \r [remember=\r as \lastr (initially 0)] in {1,...,4} {
			\draw[dashed,gray] 
				($(rank \r)!.5!(rank \lastr)$)+(0,2.7cm) -- ($(rank \r.south)!.5!(rank \lastr.south)$);
		}		
	\end{tikzpicture}
	\caption{The minimal \dfa for
	$(a+b)^\ell \setminus a^\ell$, for $\ell=4$.}
	\label{figure:sc-word-op-witness}
\end{figure}

\subsubsection{Intersection} \label{sec:intersection}
Let $L_1,L_2\subseteq\alphabet^\ell$ be two block languages, for some $\ell>0$, and their respective bitmaps~$\bm(L_1)$ and $\bm(L_2)$. The bitmap of $L_1\cap L_2$ is $\bm(L_1)\wedge\bm(L_2)$.

Let $\aut{A}_1=\auttuple{Q \cup\{\Omega_1\}}{\alphabet}{\delta_1}{q_0}{\{q_{f}\}}$ and $\aut{A}_2=\auttuple{P\cup\{\Omega_2\}}{\alphabet}{\delta_2}{p_0}{\{p_{f}\}}$ be the minimal \dfas for $L_1$ and $L_2$, respectively. As $L_1$ and $L_2$ are, in particular, both finite languages, a \dfa  $\aut{A}_3$ for $L_1\cap L_2$ has at most $mn-3(m+n)+12$ states, where $|Q|=m-1$ and $|P|=n-1$, as shown in~\cite{han08} (see Table~\ref{tab:cfin}). This bound is the result of:
\begin{enumerate}
	\item the states $(q_0,p)$, such that $p\in P\cup\{\Omega_2\}$ and $p\neq p_0$, and $(q,p_0)$, such that $q\in Q\cup\{\Omega_1\}$ and $q\neq q_0$, are not reachable from the initial state $(q_0,p_0)$, thus $m+n-2$ states can be saved;
	\item the states $(\Omega_1,p)$, such that $p\in P\setminus\{p_0\}$, and $(q,\Omega_2)$, such that $q\in Q\setminus\{q_0\}$, can be merged with $(\Omega_1,\Omega_2)$, thus $m+n-4$ states can be saved;
	\item the states $(q_f,p)$, such that $p\in P\setminus\{p_0,q_f\}$, and $(q,q_f)$, such that $q\in Q\setminus\{q_0,q_f\}$, can also be merged with $(\Omega_1,\Omega_2)$, as $(q_f, p_f)$ is the only final state, thus $m+n-6$ states can be saved.
\end{enumerate}
Thus, the number of remaining states is
$$ mn - (m+n-2) - (m+n-4) + (mn-6) = mn - 3(m+n) + 12. $$
However, for block languages more states can be saved since a state $(q, p)$ of $\aut{A}_3$ is both accessible and co-accessible if and only if $\rank(q) = \rank(p)$, for every $q\in Q$, $p\in P$. Let $Q_i$ denote the  rank $i$ in $\aut{A}_1$, and $m_i=\width(i)=|Q_i|$. Analogously, let $P_i$ denote the  rank $i$ in $\aut{A}_2$, and $n_i=\width(i)=|P_i|$. Also, we have $m=1+\sum_{i=0}^{\ell} m_i$ and $n=1+\sum_{i=0}^{\ell} n_i$, since the sink-states $\Omega_j$ do not belong to any rank, for $j\in\{1,2\}$. We have that:

\begin{lemma} \label{lemma:sc-intersection-sufficient}
	Given two minimal \dfas $\aut{A}_1$ and $\aut{A}_2$ for block languages $L_1$ and $L_2$ with block length $\ell$, respectively, a \dfa with 
	$ \sum_{i=0}^{\ell}m_in_i+1 $
	states is sufficient to recognize the intersection of $L_1$ and $L_2$, where $m_i$ and $n_i$ are the widths of rank $i$ in~$\aut{A}_1$ and $\aut{A}_2$, respectively, for $i\in [\ell]$.
\end{lemma}
\begin{proof}
	Given the above considerations, the set of states of the \dfa resulting from trimming $\aut{A}_3$ are, in the worst-case, $\bigcup_{i=0}^{\ell}Q_i\times P_i$ and a single sink-state is needed.
\end{proof}

Let us show that this bound is tight for an alphabet of fixed size, as opposed to the general case of finite languages where a growing alphabet is required~\cite{han08}. Consider the following family of languages, defined over an alphabet $\alphabet$ of size $k\geq2$, and let~$d\geq 0$ and $x\in\{0,1\}$:
	$$L_{k, d, x} = \cset{w_0\cdots w_{2d-1}\in\alphabet^{2d} \mid (\forall i\in[d-1])(i \equiv x \Mod 2 \implies w_i = w_{2d-i})}.$$
Informally, it contains the words that can be split into two halves of size $d$, where, if $x=0$ ($x=1$, resp.), then the symbols in even (odd, resp.) positions of the first half are equal to their symmetric position in the second half.

\begin{lemma} \label{lemma:sc-intersection-witness-storage}
	Let $k\geq2$, $d\geq 0$, and $x\in\{0,1\}$. Also, let $\aut{A}$ be the minimal \dfa for~$L_{k, d, x}$ over a $k$-letter alphabet $\alphabet$ and let $m_i$ be the width of $\aut{A}$, for $i\in[2d]$. Then, for $i\in[d,2d]$ we have:
	\begin{align*}
		m_i =
		\begin{cases}
			k^{\lceil \frac{2d-i}{2} \rceil},   & \text{if } x=0; \\
			k^{\lfloor \frac{2d-i}{2} \rfloor}, & \text{if } x=1, \\
		\end{cases}
	\end{align*}
	and for $i\in[d]$ we have $m_i=m_{2d-1}$.
\end{lemma}
\begin{proof}
	Let us prove for $x=0$. Refer to Figure~\ref{figure:sc-intersection-witness} for support.
	\begin{enumerate}
		\item $(\forall i\in[d,2d])(m_i=k^{\lceil \frac{2d-i}{2} \rceil})$: \\
			Let $w_1,w_2\in\alphabet^{2d-i}$ such that they differ at least in one even position. Now, let $w_3 = \letter^{2(i-d)}\reversal{w_1}$, for some $\letter\in\alphabet$. It is easy to see that $w_1w_3\in L_{k, d, 0}$ but $w_2w_3\notin L_{k, d, 0}$, so $w_1$ and $w_2$ have different quotients, and so they have to reach different states. Therefore, the number of states on rank $i$ of $\aut{A}$ is given by~$k^{\lceil \frac{2d-i}{2} \rceil}$, where the exponent is the number of odd integers between $i$ and~$2d-1$.
		\item $(\forall i\in[d])( m_i=m_{2d-i})$: \\
			It is easy to see that the proposition holds, as $L_{k, d, x}=\reversal{L_{k, d, x}}$, following a similar argument to the one provided in the proof for Lemma~\ref{lemma:nsc-nfa}.
	\end{enumerate}
	For $x=1$, the exponent in $k^{\lfloor \frac{2d-i}{2} \rfloor}$ is the number of even integers between $i$ and~$2d-1$, so the proof is similar.
\end{proof}

\begin{figure}[ht]
	\centering
	\def\x{1cm}
	\begin{tikzpicture}[>=stealth', shorten >=1pt, auto, node distance=2cm, initial text={}, scale=0.6, every node/.style={scale=0.8}]
		\node[sstate,initial]                                      (0)  {$q_0$};
		\node[sstate,above of=0,right of=0,yshift=-\x]             (1)  {$q_1$};
		\node[sstate,below of=0,right of=0,yshift=\x]              (2)  {$q_2$};
		\node[sstate,right of=1]                                   (3)  {$q_3$};
		\node[sstate,right of=2]                                   (4)  {$q_4$};
		\node[sstate,above of=3,right of=3,yshift=-.5\x]           (5)  {$q_5$};
		\node[sstate,right of=3]                                   (6)  {$q_6$};
		\node[sstate,right of=4]                                   (7)  {$q_7$};
		\node[sstate,below of=4,right of=4,yshift=.5\x]            (8)  {$q_8$};
		\node[sstate,right of=6]                                   (9)  {$q_9$};
		\node[sstate,right of=7]                                   (10) {$q_{10}$};
		\node[sstate,right of=9]                                   (11) {$q_{11}$};
		\node[sstate,right of=10]                                  (12) {$q_{12}$};
		\node[sstate,accepting,right of=11, below of=11,yshift=\x] (13) {$q_{13}$};

		\path[->]
			(0)  edge node               {$a$}   (1)
				 edge node[swap]         {$b$}   (2)
			(1)  edge node               {$a,b$} (3)
			(2)  edge node               {$a,b$} (4)
			(3)  edge node               {$a$}   (5)
			(3)  edge node[pos=.3]       {$b$}   (6)
			(4)  edge node[pos=.3]       {$b$}   (7)
			(4)  edge node[swap]         {$a$}   (8)
			(5)  edge node               {$a$}   (9)
			(8)  edge node[swap,pos=.5]  {$a$}   (10)
			(6)  edge node[pos=.3]       {$b$}   (9)
			(7)  edge node[pos=.3]       {$b$}   (10)
			(9)  edge node               {$a,b$} (11)
			(10) edge node               {$a,b$} (12)
			(11) edge node[pos=.15]      {$a$}   (13)
			(12) edge node[swap,pos=.15] {$b$}   (13)
		;
			
		\path[gray]
				(8)++(-90:5ex) node (rank 3) {rank $3$}
				(rank 3 -| 10) node (rank 2) {rank $2$}
				(rank 3 -| 12) node (rank 1) {rank $1$}
				(rank 3 -| 13) node (rank 0) {rank $0$}
				(rank 3 -| 4)  node (rank 4) {rank $4$}
				(rank 3 -| 2)  node (rank 5) {rank $5$}
				(rank 3 -| 0)  node (rank 6) {rank $6$}
		;
		\foreach \r [remember=\r as \lastr (initially 0)] in {1,...,6} {
			\draw[dashed,gray] 
				($(rank \r)!.5!(rank \lastr)$)+(0,8.6cm) -- ($(rank \r.south)!.5!(rank \lastr.south)$);
		}		
	\end{tikzpicture}
	\caption{The minimal \dfa for $L_{k,d,x}$ with $k=2$, $d=3$ and $x=0$.}
	\label{figure:sc-intersection-witness}
\end{figure}

\begin{lemma}\label{lemma:sc-intersection-necessary}
	For some $d\geq0$ and $k\geq2$, let $\aut{A}_1$ and $\aut{A}_2$ be \dfas that accept $L_{k, d, 0}$ and $L_{k, d, 1}$, respectively. A \dfa that recognizes $L_{k, d, 0}\cap L_{k, d, 1}$ needs at least $\sum_{i=0}^{2d}m_in_i+1$ states, where $m_i$ and $n_i$ correspond to the widths of rank $i\in[2d]$ in $\aut{A}_1$ and $\aut{A}_2$, respectively.
\end{lemma}
\begin{proof}
	It is easy to see that
	$$L_{k,d,0} \cap L_{k,d,1} = \{\, w\reversal{w} \mid w\in\alphabet^d \,\} = L_{k,d}$$
	is the family of languages provided as witness for the maximal nondeterministic state complexity of a block language in Section~\ref{sec:sc-nfa}. In particular, we concluded that a minimal NFA for $L_{k,d}$ must be of maximal size (Lemma~\ref{lemma:nsc-nfa}), and that the NFA must be deterministic (Corollary~\ref{corollary:nsc-nfa}). Let $\aut{A}_3$ be that \dfa with set of states $S=S_0\cup S_1\cup\ldots\cup S_{2d}$, such that $S_i$ corresponds to the set of states in rank $i\in[2d]$ of $\aut{A}_3$. As $\aut{A}_3$ is an NFA of maximal size, $|S_i|=|S_{2d-i}|=k^{2d-i}$, and, as seen in the proof of Lemma~\ref{lemma:sc-intersection-witness-storage}, we have $m_i=m_{2d-i}=k^{\lceil \frac{2d-i}{2} \rceil}$ and $n_i=n_{2d-i}=k^{\lfloor \frac{2d-i}{2} \rfloor}$, for $i\in[d, 2d]$. In fact,
	$$|S_i| = m_in_i = k^{\lceil \frac{2d-i}{2} \rceil}k^{\lfloor \frac{2d-i}{2} \rfloor}=k^{2d-i}\text,$$
	as desired.
\end{proof}

From Lemmas~\ref{lemma:sc-intersection-sufficient} and~\ref{lemma:sc-intersection-necessary} we have:
\begin{theorem}\label{theorem:sc-intersection}
	Given two block languages $L_1,L_2\subseteq\alphabet^\ell$, for $\ell>0$, with minimal \dfas~$\aut{A}_1$ and $\aut{A}_2$, respectively, we have
		$$ \dsc(L_1\cap L_2) \leq \sum_{i=0}^{\ell}m_in_i+1, $$
		and the bound is tight for alphabets of size at least 2, where $m_i$ and $n_i$ are the widths of rank $i$ in $\aut{A}_1$ and $\aut{A}_2$, respectively, for $i\in [\ell]$.
\end{theorem}

For the nondeterministic state complexity, the bounds are the same except that the sink-state is not considered. In fact, the family witness languages  for the tightness of deterministic state complexity is also a witness for the nondeterministic one.

\begin{theorem}\label{theorem:nsc-intersection}
	Let $\aut{A}_1$ be an $m$-state minimal \nfa  for a block language $L_1\subseteq\alphabet^\ell$, and $\aut{A}_2$ be an $n$-state minimal \nfa for $L_2\subseteq\alphabet^\ell$, for some $\ell >0$. 
Also, let $m_i$ and~$n_i$ be the widths of rank $i$ of $\aut{A}_1$ and $\aut{A}_2$, respectively. Then, an \nfa  with $\sum_{i=0}^{\ell}m_in_i$ states is sufficient to recognize $L_1\cap L_2$ and the bound is tight for $k>1$.
\end{theorem}
\begin{proof} 
	That $\sum_{i=0}^{\ell}m_in_i$ states are sufficient follows from the previous discussions. That are necessary results from considering again the languages $L_{k, d, x}$, for $k>1$, $d>0$ and $x\in\{0,1\}$. Similar arguments to the ones used for Corollary~\ref{corollary:nsc-nfa} can be given to justify that the minimal \nfas for $L_{k, d, x}$ are, in fact, deterministic. Then, we have $\dsc(L_{k, d, x})-1 =\nsc(L_{k, d, x})$. The remaining of the proof follows similarly to the proof of Lemma~\ref{lemma:sc-intersection-necessary}.
\end{proof}

\subsubsection{Union} \label{sec:union}
Let $L_1,L_2\subseteq\alphabet^\ell$ be two block languages, for some $\ell>0$ and their respective bitmaps~$\bm(L_1)$ and $\bm(L_2)$. The bitmap of $L_1\cup L_2$ is $\bm(L_1)\vee\bm(L_2)$.

Let $\aut{A}_1=\auttuple{Q \cup\{\Omega_1\}}{\alphabet}{\delta_1}{q_0}{\{q_{f}\}}$ and $\aut{A}_2=\auttuple{P\cup\{\Omega_2\}}{\alphabet}{\delta_2}{p_0}{\{p_{f}\}}$ be the minimal \dfas for $L_1$ and $L_2$, respectively, with $|Q|=m-1$ and $|P|=n-1$. Again, as $L_1$ and $L_2$ are both finite languages, a DFA $\aut{A}_3$ for $L_1\cup L_2$ has at most $mn-(m+n)$ states, as shown in~\cite{han08} (see Table~\ref{tab:cfin}). This bound is the result of:
\begin{enumerate}
	\item the states $(q_0,p)$, such that $p\in P\cup\{\Omega_2\}$ and $p\neq p_0$, and $(q,p_0)$, such that $q\in Q\cup\{\Omega_1\}$ and $q\neq q_0$, are not reachable from the initial state $(q_0,p_0)$, thus $m+n-2$ states can be saved;
	\item the final states $(q_{f},\Omega_2)$, $(\Omega_1, p_{f})$, and $(q_f, p_f)$ can be merged into a single final state, so $2$ extra states can be saved.
\end{enumerate}
However, again, for block languages one only needs to consider pairs of states $(q,p)$ such that $\rank(q) = \rank(p)$, for $q\in Q, p\in P$. Let $Q_i$ denote the rank $i$ in~$\aut{A}_1$, and $m_i=\width(i)=|Q_i|$. Analogously, let $P_i$ denote the rank $i$ in $\aut{A}_2$, and $n_i=\width(i)=|P_i|$. Also, we have $m=1+\sum_{i=0}^{\ell} m_i$ and $n=1+\sum_{i=0}^{\ell} n_i$, since the sink-states $\Omega_j$ do not belong to any rank, for $j\in\{1,2\}$. We have that:
	
\begin{lemma} \label{lemma:sc-union-sufficient}
	Given two \dfas $\aut{A}_1$ and $\aut{A}_2$ for block languages $L_1$ and $L_2$ with block length $\ell$, respectively, a \dfa with 
	$\sum_{i=1}^{\ell-1}(m_in_i+m_i+n_i)+3$
	states is sufficient to recognize the union of $L_1$ and $L_2$, where $m_i$ and $n_i$ are the widths of rank $i$ in $\aut{A}_1$ and~$\aut{A}_2$, respectively, for $i\in [1,\ell-1]$.
\end{lemma}
\begin{proof}
	Let $\aut{A}_3$ be the product automaton from $\aut{A}_1$ and $\aut{A}_2$. As mentioned above, the final states $(q_{f}, \Omega_2)$ and $(\Omega_1, p_{f})$ can be merged with $(q_{f}, p_{f})$, and a state $(p,q)\in Q\times P$ is only accessible from the initial state if $\rank(p)=\rank(q)$. Therefore, the \dfa resulting from trimming $\aut{A}_3$ has a single initial state, a final state and a sink-state, and also the states $(Q_i\times P_i) \cup (Q_i\times\{\Omega_2\}) \cup (\{\Omega_1\}\times P_i)$, at each rank $i\in[1, \ell-1]$. Thus, the sufficient  number of states follows.
\end{proof}

In fact, the bound is tight for an alphabet with size at least $3$.
\begin{lemma} \label{lemma:sc-union-necessary}
	Given two \dfas $\aut{A}_1$ and $\aut{A}_2$ for block languages $L_1\subseteq\alphabet^\ell $ and $L_2\subseteq\alphabet^\ell$, respectively,
	the necessary number of states of a \dfa recognizing the union of $L_1$ and $L_2$
	is $\sum_{i=1}^{\ell-1}(m_in_i+m_i+n_i)+3$,
	where $m_i$ and $n_i$ are the widths of rank $i$ in $\aut{A}_1$ and $\aut{A}_2$, respectively, for $i\in [1,\ell-1]$ and $|\alphabet|>2$.
\end{lemma}
\begin{proof}
	Since $\aut{A}_1$ and $\aut{A}_2$ are deterministic, $m_{\ell-1}$ and $n_{\ell-1}$, the $\width(\ell-1)$ in $\aut{A}_1$ and~$\aut{A}_2$, respectively, is bounded by $k=|\alphabet|$ and not equal to $0$. Analogously, $\width(\ell-1)$ of the \dfa for the union of $\lang{\aut{A}_1}$ and $\lang{\aut{A}_2}$ is also at most $k$. When~$k=2$, it is easy to see that the inequality $0< m_{\ell-1}n_{\ell-1}+m_{\ell-1}+n_{\ell-1}\leq k$ has no solutions.
	
	Now, consider the languages $L_{1,\ell}=(a+c)^\ell$ and $L_{2,\ell}=(b+c)^\ell$, and let $\aut{A}_1$ and $\aut{A}_2$ be the minimal \dfas that recognize them, respectively, for some $\ell>0$ and $\alphabet=\{a,b,c\}$. It is easy to see that $\dsc(L_{1,\ell})=\dsc(L_{2,\ell})=\ell+2$ and $m_i=n_i=1$, for every $i\in[\ell]$. The minimal \dfa that recognizes the language $L_{1,\ell}\cup L_{2,\ell}$ requires $3$ states at each rank~$i\in[1,\ell-1]$: one state when an $a$ has already been read, so the word is in $L_{1,\ell}$; one state when a $b$ has already been read, so the word is in $L_{2,\ell}$; and one state  when only $c$'s have been read (refer to Figure~\ref{figure:sc-union-witness}). Then,
		$ \dsc(L_{1,\ell}\cup L_{2,\ell}) = \sum_{i=1}^{\ell-1}(n_im_i+n_i+m_i)+3 = 3\ell.$
\end{proof}

\begin{figure}[ht]
	\centering
	\begin{tikzpicture}[>=stealth', shorten >=1pt, auto, node distance=2cm, initial text={},every node/.style={scale=0.8}]
		\node[sstate,initial]                            (0)  {$q_0$};
		\node[sstate,above of=0,right of=0]              (1)  {$q_1$};
		\node[sstate,right of=0]                         (2)  {$q_2$};
		\node[sstate,below of=0,right of=0]              (3)  {$q_3$};
		\node[sstate,above of=2,right of=2,yshift=.3cm]
		  (4)  {$q_4$};
		\node[sstate,right of=2]                         (5)  {$q_5$};
		\node[sstate,below of=2,right of=2,yshift=-.3cm] (6)  {$q_6$};
		\node[sstate,above of=5,right of=5]              (7)  {$q_7$};
		\node[sstate,right of=5]                         (8)  {$q_8$};
		\node[sstate,below of=5,right of=5]              (9)  {$q_9$};
		\node[sstate,right of=8,accepting]               (10) {$q_{10}$};

		\path[->]
			(0) edge node                            {$a$}     (1)
				edge node[pos=.3]                    {$c$}     (2)
				edge node[swap]                      {$b$}     (3)
			(1) edge node[inner sep=.5,pos=.7]       {$a,c$}   (4)
			(4) edge node[inner sep=.5,pos=.25]      {$a,c$}   (7)
			(7) edge node                            {$a,c$}   (10)
			(3) edge node[swap,inner sep=.5,pos=.7]  {$b,c$}   (6)
			(6) edge node[swap,inner sep=.5,pos=.25] {$b,c$}   (9)
			(9) edge node[swap]                      {$b,c$}   (10)
			(2) edge node                            {$a$}     (4)
			    edge node[pos=.3]                    {$c$}     (5)
			    edge node[swap]                      {$b$}     (6)
			(5) edge node                            {$a$}     (7)
			    edge node[pos=.3]                    {$c$}     (8)
			    edge node[swap]                      {$b$}     (9)
			(8) edge node                            {$a,b,c$} (10)
			;
			
		\path[gray]
				(6)++(-90:5ex) node (rank 2) {rank $2$}
				(rank 2 -| 3)  node (rank 3) {rank $3$}
				(rank 2 -| 0)  node (rank 4) {rank $4$}
				(rank 2 -| 9)  node (rank 1) {rank $1$}
				(rank 2 -| 10) node (rank 0) {rank $0$}
		;
		\foreach \r [remember=\r as \lastr (initially 0)] in {1,...,4} {
			\draw[dashed,gray] 
				($(rank \r)!.5!(rank \lastr)$)+(0,4.2cm) -- ($(rank \r.south)!.5!(rank \lastr.south)$);
		}		
	\end{tikzpicture}
	\caption{The minimal \dfa for $L_{1,\ell}\cup L_{2,\ell}=(a+c)^\ell+(b+c)^\ell$, with $\ell=4$.}
	\label{figure:sc-union-witness}
\end{figure}

From Lemmas~\ref{lemma:sc-union-sufficient} and~\ref{lemma:sc-union-necessary}, we have:
\begin{theorem}\label{theorem:sc-union}
		Given two block languages $L_1,L_2\subseteq\alphabet^\ell$, for $\ell>0$, with minimal \dfas $\aut{A}_1$ and $\aut{A}_2$, respectively, we have
		$$ \dsc(L_1\cup L_2) \leq \sum_{i=1}^{\ell-1}(m_in_i+m_i+n_i)+3, $$
		and the bound is tight for $|\alphabet|\geq 3$, where $m_i$ and $n_i$ are the widths of rank $i$ in $\aut{A}_1$ and $\aut{A}_2$, respectively, for $i\in [1,\ell-1]$.
\end{theorem}

For the nondeterministic state complexity, the upper bound is the same as for finite languages over the same alphabet.
\begin{theorem}	\label{theorem:nsc-union}
	Let $L_1, L_2\subseteq\alphabet^\ell$ with $\ell>0$ and $|\alphabet|=k$, such that $\nsc(L_1)=m$ and $\nsc(L_2)=n$. Then, $\nsc(L_1\cup L_2)\leq m+n-2$, and this bound is reached.
\end{theorem}
\begin{proof}
	Let $\aut{A}_1$ and $\aut{A}_2$ be minimal \nfas for $L_1$ and $L_2$ with $m$ and $n$ states, respectively. An \nfa for $L_1\cup L_2$ can be constructed by merging the initial states of $\aut{A}_1$ and $\aut{A}_2$, as well as the final states, so $2$ states can be saved. In fact, this bound is tight, as the following families of languages suggest. Let $L_{1,\ell}=\{a^\ell\}$ and $L_{2,\ell}=\{b^\ell\}$, for some $\ell>0$ and $\alphabet=\{a,b\}$. We have that $\nsc(L_{1,\ell})=\nsc(L_{2,\ell})=\ell+1$ and $\nsc(L_{1,\ell}\cup L_{2,\ell}) = 2\ell$.
\end{proof}

\subsubsection{Concatenation} \label{sec:concatenation}
Consider two languages $L_1 \subseteq \alphabet^{\ell_1}$ and $L_2 \subseteq \alphabet^{\ell_2}$, for some $\ell_1, \ell_2 > 0$ and $|\alphabet|=k$, with bitmaps $\bm(L_1)$ and $\bm(L_2)$, respectively. The bitmap for the language $L_1L_2$ is obtained by replacing each $1$ in $\bm(L_1)$ by $\bm(L_2)$ and each $0$ by $0^{k^{\ell_2}}$. This ensures that each word of $L_1L_2$ is obtained by concatenating a word of-$L_1$ with a word of $L_2$ and for each word obtained in such way the correspondent bit in $\bm(L_1L_2)$ is set to $1$.

The deterministic state complexity of the concatenation for block languages coincides with the one for the finite languages when the first operand, $L_1$, has a single final state in its minimal \dfa. Therefore, we have the following exact upper bound.

\begin{theorem}\label{theorem:sc-concatenation}
Let $L_1\subseteq\alphabet^{\ell_1}$ and $L_2\subseteq \alphabet^{\ell_2}$, for some $\ell_1,\ell_2 > 0$, be two block languages over a $k$-letter alphabet, where $\dsc(L_1)=m$ and $\dsc(L_2)=n$. Then, $\dsc(L_1L_2) = m + n - 2$.
\end{theorem}
\begin{proof}
	Let $\aut{A}_1$ and $\aut{A}_2$ be the minimal \dfas for $L_1$ and $L_2$, respectively. Also, let~$\aut{A}_3$ be the minimal \dfa for $L_1L_2$. Considering the bitmaps for these languages, the width of the rank $i$ of $\aut{A}_3$ is $|\bmsetl{L_2}{i}|$, if $i\in[\ell_2]$, or is $|\bmsetl{L_1}{i-\ell_2}|$, if $i\in[\ell_2+1,\ell_1+\ell_2]$. Then, $\aut{A}_3$ saves $2$ states by reusing the final state of $\aut{A}_1$ for the initial state of $\aut{A}_2$ (alternatively, reusing the initial state of $\aut{A}_2$ for the final state of $\aut{A}_1$) and also by eliminating one of the sink-states.
\end{proof}

For the nondeterministic state complexity, the same result is expected, coinciding with the nondeterministic state complexity for the finite languages.
\begin{theorem}\label{theorem:nsc-concatenation}
Let $L_1\subseteq \alphabet^{\ell_1}$ and $L_2\subseteq \alphabet^{\ell_2}$, for some $\ell_1, \ell_2 > 0$, be two block languages over a $k$-letter alphabet, where $\nsc(L_1)=m$ and $\nsc(L_2)=n$. Then, $\nsc(L_1L_2) = m + n - 1$.	
\end{theorem}

In fact, any two languages $L_1 \subseteq \alphabet^{\ell_1}$ and $L_2 \subseteq \alphabet^{\ell_2}$ result in a family of witness languages, as this operation preserves the ranks of the automata of the operands, both in the deterministic and nondeterministic case. 
\begin{example}\label{example:concatenation}
	Let $L_{1,\ell_1} = \{a^{\ell_1}\}$ and $L_{2,\ell_2} = \{a^{\ell_2}\}$, for $\ell_1, \ell_2>0$ and $\alphabet=\{a\}$. We have that $\dsc(L_{1,\ell_1})=\ell_1+2$, $\dsc(L_2)=\ell_2+2$, and $\dsc(L_{1,\ell_1}L_{2,\ell_2})=\ell_1+\ell_2+2$. We also have $\nsc(L_{1,\ell_1})=\ell_1+1$, $\nsc(L_{2,\ell_2})=\ell_2+1$, and $\nsc(L_{1,\ell_1}L_{2,\ell_2})=\ell_1+\ell_2+1$.
\end{example}

\subsubsection{Block Complement} \label{sec:bcomplement}
Consider a language $L\subseteq\alphabet^\ell$, for some $\ell>0$ and alphabet of size $k>1$, and let~$\bm$ be its bitmap. We now consider the block complement language, namely $\alphabet^\ell\setminus L$, which we shall denote by $\complement{L}^\ell$. The bitmap of the language $\complement{L}^\ell$, namely $\complement{\bm}$, is given by complementing every bit of $\bm$.

\begin{theorem} \label{theorem:sc-bcomplement}
	Let $L\subseteq\alphabet^\ell$, with $\ell>0$, be a block language with $|\alphabet|=k$, such that $\dsc(L)=m$. Then, $m-(\ell-1)\leq \dsc(\complement{L}^\ell)\leq m+(\ell-1)$.
\end{theorem}
\begin{proof}
	The number of states on a rank $i\in[\ell]$ of the minimal \dfa for $\complement{L}^\ell$ is given by the cardinality of $\complement{\bmset{}}_i$, the set of the non-null factors of length $k^i$ on the bitmap $\complement{\bm}$. If, for some $j\in[k^{\ell-i}-1]$, we have that $\bmsegment{i}{j} = 0\cdots0$, which by definition implies that $\bmsegment{i}{j}\notin\bmset{i}$, then $\complement{\bmsegment{i}{j}}=1\cdots1$ and so $\overline{\bmsegment{i}{j}}\in\overline{\bmset{}}_i$. Moreover, the complement may also occur. If $\bmsegment{i}{j}$ contains both zeros and ones, then both $\bmsegment{i}{j}\in\bmset{i}$ and $\complement{\bmsegment{i}{j}}\in\complement{\bmset{}}_i$, so the operation does not alter the cardinality of the set. Therefore, $\big||\bmset{i}|-|\complement{\bmset{}}_i|\big|\leq 1$, and so the upper bound follows.
	
	Let $L_\ell=\{a^\ell\}$, for $\ell>0$. As we previously saw in Theorem~\ref{theorem:sc-word-op-witness}, $\dsc(L)=\ell+2$, and $\dsc(\complement{L}^\ell) = 2\ell+1$.
\end{proof}

For the nondeterministic state complexity of the block complement operation, we have that the bound meets the one of the complement from the general case for finite languages considering the determinization cost of block languages. Also, this bound is asymptotically tight for alphabets of size at least $2$.

\begin{lemma}\label{lemma:nsc-bcomplement-sufficient} 
	Let $L\subseteq\alphabet^\ell$ be a block language with $|\alphabet|=k$, such that $L$ is accepted by an $m$-state \nfa. Then, $2^{O(\sqrt{m})}$ states are sufficient for an \nfa to recognizes $\complement{L}^\ell$.
\end{lemma}
\begin{proof}
	Let $\aut{A}_1$ be an \nfa for $L$ with $m$ states. The minimal \dfa~$\aut{A}_2$ for $L$ has at most $2^{O(\sqrt{m})}$ states~\cite{KarhOkho:2O14}. Furthermore, the minimal \dfa~$\aut{A}_3$ for $\complement{L}^\ell$ has at most $\ell+1$ more states then $\aut{A}_2$, as shown in Theorem~\ref{theorem:sc-complement}. Of course, the nondeterministic state complexity is bounded by the deterministic state complexity, that is, $\nsc(L)\leq\dsc(L)$, so the sufficient number of states follows.
\end{proof}

Consider the following family presented by Karhumäki and Okhotin~\cite{KarhOkho:2O14}:
$$ L_{k, d} = \cset{w_0\cdots w_{2d-1} \mid (\exists i\in[d-1])( w_i=w_{i+d}\in\alphabet\setminus\{\letter_{k-1}\})} $$ 
defined over a $k$-ary alphabet $\alphabet = \{\letter_0,\ldots,\letter_{k-1}\}$. Informally, this language contains words that can be split into two halves of size $d$, such that there is at least one position in the first half that matches its counterpart in the second one, and it is different than the \emph{prohibited symbol} $\letter_{k-1}$. 

\begin{proposition}[\cite{KarhOkho:2O14}] \label{proposition:nsc-complement}
	For each $k\geq2$ and $d\geq2$, the language $L_{k,d}$ is recognized by an \nfa with $(k-1)d^2+2d$ states.
\end{proposition}

Let us now focus on the block complement of these languages.

\begin{lemma} \label{lemma:nsc-bcomplement}
	For each $k\geq2$ and $d\geq2$, the language $\complement{L}_{k,d}^{2d}$, defined over a $k$-letter alphabet $\alphabet=\{\letter_0,\ldots,\letter_{k-1}\}$, requires at least $k^d$ states on the $d$-th rank.
\end{lemma}
\begin{proof}
	First, notice that $\complement{L}_{k,d}^{2d}$, the block complement of the language $L_{k,d}$, can be formally defined as
	$$ \complement{L}_{k,d}^{2d} = \cset{w_0\cdots w_{2d-1} \mid (\forall i\in[d-1])( w_i\neq w_{i+d} \text{ or } w_i=\letter_{k-1})}.$$
	Let $u$ and $v$ be two words in $\alphabet^d$ such that $\letter'$ and $\letter''$ are the $i$-th symbols of $w_1$ and $w_2$, respectively, with $\letter'\neq \letter''$ and $i\in[d]$. If $\letter'=\letter_{k-1}$ then, considering $w_3=\letter_{k-1}^{i-1}\,\letter''\,\letter_{k-1}^{d-i}$, we have $w_1w_3\in\overline{L}_{k,d}^{2d}$ but $w_2w_3\notin\overline{L}_{k,d}^{2d}$. If $\letter''=\letter_{k-1}$ then, taking $w_3=\letter_{k-1}^{i-1}\,\letter'\,\letter_{k-1}^{d-i}$, we have $w_1w_3\notin\overline{L}_{k,d}^{2d}$ but $w_2w_3\in\overline{L}_{k,d}^{2d}$. As a consequence, $w_1^{-1}\overline{L}_{k,d}^{2d} \neq w_2^{-1}\overline{L}_{k,d}^{2d}$. Therefore, one state in rank $d$ is needed for each word in $\alphabet^d$.
\end{proof}

With these results, it is possible to determine that the nondeterministic state complexity for the block complement operation given in Lemma~\ref{lemma:nsc-bcomplement-sufficient} is tight.

\begin{theorem}\label{theorem:nsc-bcomplement}
	Let $m\geq2$ and $\alphabet$ an alphabet of size $k\geq2$. Then, there exists a language $L\subseteq\alphabet^\ell$, for some $\ell>0$, such that $\nsc(L)=m$ and $\nsc(\complement{L}^\ell)\in2^{\Omega\left(\sqrt{m}\right)}$. 
\end{theorem}
\begin{proof}
	Consider $d$ as the largest integer for which $(k-1)d^2+2d\leq m$. As shown in~\cite{KarhOkho:2O14}, we have that
	$$ d =  \left\lfloor \sqrt{\frac{m}{k-1} + \frac{1}{(k-1)^2}} - \frac{1}{k-1} \right\rfloor \geq \sqrt{\frac{m}{k-1}}-2. $$
	Then, $L_{k,d}$ is a language recognized by an-$m$-state \nfa, while every \nfa for $\complement{L}_{k,d}^\ell$ requires, by Lemma~\ref{lemma:nsc-bcomplement}, at least
	$$k^d = k^{\left\lfloor \sqrt{\frac{m}{k-1} + \frac{1}{(k-1)^2}} - \frac{1}{k-1} \right\rfloor} \geq k^{\sqrt{\frac{m}{k-1}}-2} \in2^{\Omega(\sqrt{m})}$$
	states, as desired.
\end{proof}

In the next sections, we consider some operations that are not closed for block languages.

\subsubsection{Kleene Star and Plus} \label{sec:kleene}
Let $L\subseteq\alphabet^\ell$, for some $\ell>0$ and $\bm$ its bitmap. From $\bm$ one can obtain the minimal \dfa for $L$, namely $\aut{A}_1=\langle Q,\alphabet,\delta_1,q_0,\{q_f\}\rangle$.  A \dfa~$\aut{A}_2=\langle Q\setminus \{q_f\},\alphabet,\delta_2,q_0,\{q_0\}\rangle$ recognizes the language $\star{L}$ if:
\begin{enumerate}
	\item $\delta_2(q,\letter)=q_0$, for all $q\in Q$ and $\letter\in\alphabet$, such that $\delta_1(q,\letter)=q_f$;
	\item $\delta_2(q,\letter)=\delta_1(q,\letter)$, for all the remaining pairs $(q,\letter)\in Q\times\alphabet$.
\end{enumerate}
That is, the \dfa for $\star{L}$ is given by substituting  all the transitions with final state as the target state to transitions to the initial state. The same applies for the \nfa for~$\star{L}$, as the following theorem states.
\begin{theorem}\label{theorem:dnsc-kleene}
	Let $L\subseteq\alphabet^\ell$, with $\ell>0$, be a block language with $\dsc(L) = n$ and $\nsc(L) = m$. Then, $\dsc(\star{L}) = n-1$ and $\nsc(\star{L}) = m-1$.
\end{theorem}

Moreover, a \dfa~$\aut{A}_3=\auttuple{Q}{\alphabet}{\delta_3}{q_0}{\{q_f\}}$ recognizes the language $\plus{L}$ if 
\begin{enumerate}
	\item $\delta_3(q_f,\letter)= \delta_1(q_0,\letter)$, for all $\letter\in\alphabet$;
	\item $\delta_3(q,\letter)=\delta_1(q,\letter)$, for all the remaining pairs $(q,\letter)\in Q\times\alphabet$.
\end{enumerate}
Again, the \nfa for $\plus{L}$ is given by applying the same changes to the minimal \nfa for $L$. 

\begin{theorem}\label{theorem:dnsc-plus}
	Let $L\subseteq\alphabet^\ell$, with $\ell>0$. Then, $\dsc(\plus{L}) = \dsc(L)$ and $\nsc(\plus{L}) =\nsc(L)$.
\end{theorem}

\subsubsection{Block Language Stencil} \label{section:block-extension}
The \emph{block language stencil} operation builds upon the concept of  \emph{cover automata} for a finite language $L$. Those automata accept the  words in $L$ as well as some additional words that are longer than any word in $L$~\cite{campeanu99}. For a block language $L\subseteq \alphabet^\ell$, where~$\ell>0$, we are interested  in constructing an automaton that accepts $L$ but also all words of lengths other than $\ell$, i.e., the language $L\cup \alphabet^{\not=\ell}$. We denote this language by $\extended{L}$. 

Given a minimal \dfa~$\aut{A}_1=\langle Q\cup\{\Omega\},\alphabet,\delta_1,q_0,\{q_f\}\rangle$ recognizing a block language $L\subseteq \alphabet^\ell$, with $\ell>0$, we construct a  \dfa~$\aut{A}_2=\langle Q\cup P,\alphabet,\delta_2,q_0,Q\cup P\setminus\{p_0\}\rangle$, with $P=\{p_0,\ldots,p_{\ell-1}\}$ and $Q\cap P= \emptyset$, and the transition function defined as follows:
\begin{enumerate}
 	\item $\delta_2(q,\letter) = p_{\rank(q)-1}$, 
 	for all $(q,\letter)\in Q\times\alphabet$ such that 
 	 $\delta_1(q,\letter)=\Omega$; \label{item:be-item-1}
	\item $\delta_2(p_i,\letter)=p_{i-1}$, for all $i\in[1,\ell-1]$ and $\letter\in \alphabet$;
	\item $\delta_2(p_0,\letter)=q_f$, for  all $\letter \in \alphabet$;
	\item $\delta_2(q,\letter)=\delta_1(q,\letter)$, for all the remaining pairs $(q,\letter)\in Q\times\alphabet$.
\end{enumerate}
Then, we have
\begin{lemma} \label{lemma:minimal-dfa-extension}
	The automaton $\aut{A}_2$ is minimal and recognizes $\extended{L}$.
\end{lemma}

\begin{theorem} \label{theorem:sc-extension}
	Let $L\subseteq\alphabet^\ell$, with $\ell>0$, be a block language with $\dsc(L) = m$ and $\nsc(L)=n$. Then, $\dsc(\extended{L}) = m+\ell-1$ and $\nsc(\extended{L})=n+\ell$.
\end{theorem}
\begin{proof}
	The fact that it is sufficient follows from the construction discussed above. To show that it is tight, consider $L_\ell=\{a^\ell\}$ over the alphabet $\alphabet=\{a,b\}$, for $\ell>0$. We have $\dsc(L_\ell)=\nsc(L_\ell)+1=\ell+2$, while $\dsc(\extended{\{a^\ell\}})=\nsc(\extended{\{a^\ell\}})=2\ell+1$. Figure~\ref{figure:sc-extension-witness} presents the minimal automaton for $\ell=3$.
\end{proof}

\begin{figure}[ht]
	\centering
	\def\shift{1cm}
	\begin{tikzpicture}[>=stealth', shorten >=1pt, auto, node distance=2cm, initial text={},every node/.style={scale=0.8}]
		\node[sstate,initial,accepting]                              (0)    {$q_0$};
		\node[sstate,above of=0,right of=0,yshift=-\shift,accepting] (1)    {$q_1$};
		\node[sstate,below of=0,right of=0,yshift=\shift,accepting]  (2)    {$p_2$};
		\node[sstate,right of=1,accepting]                           (3)    {$q_2$};
		\node[sstate,right of=2,accepting]                           (4)    {$p_1$};
		\node[sstate,right of=3,accepting]                           (5)    {$q_3$};
		\node[sstate,right of=4]                                     (6)    {$p_0$};

		\path[->]
			(0) edge                node                       {$a$}   (1)
				edge                node[swap]                 {$b$}   (2)
			(2) edge                node[swap]                 {$a,b$} (4)
			(4) edge                node[swap]                 {$a,b$} (6)
			(6) edge                node[swap,inner sep=.55]   {$a,b$} (5)
			(1) edge                node[pos=.3]               {$a$}   (3)
				edge                node[inner sep=.5,pos=.25] {$b$}   (4)
			(3) edge                node[pos=.3]               {$a$}   (5)
				edge                node[inner sep=.5,pos=.25] {$b$}   (6)
			(5) edge[loop right] node                          {$a,b$} (5)
			;
			
		\path[gray]
				(4)++(-90:5ex) node (rank 1) {rank $1$}
				(rank 1 -| 2)  node (rank 2) {rank $2$}
				(rank 1 -| 0)  node (rank 3) {rank $3$}
				(rank 1 -| 6)  node (rank 0) {rank $0$}
		;
		\foreach \r [remember=\r as \lastr (initially 0)] in {1,...,3} {
			\draw[dashed,gray] 
				($(rank \r)!.5!(rank \lastr)$)+(0,2.8cm) -- ($(rank \r.south)!.5!(rank \lastr.south)$);
		}		
	\end{tikzpicture}
	\caption{The minimal \dfa (a minimal NFA) for $\extended{\{a^\ell\}}$, for $\ell=3$ and $\alphabet=\{a,b\}$. 
	 The state $q_3$ was the final state for the minimal \dfa of $\{a^3\}$.
	}
	\label{figure:sc-extension-witness}
\end{figure}

\subsubsection{Complement} \label{sec:complement}
The complement of a block language $L\subseteq\alphabet^\ell$, for some $\ell>0$, $\complement{L}=\star{\alphabet}\setminus L$, is exactly  $\complement{L}^\ell\cup \alphabet^{\not=\ell}$, i.e., $\extended{(\complement{L}^\ell)}$.

For the deterministic state complexity, the bounds match with those for general regular languages, where it suffices to swap final and non-final states.
\begin{theorem}\label{theorem:sc-complement}
	Let $L\subseteq\alphabet^\ell$ be a block language. Then, $\dsc(L)=\dsc(\complement{L})$.
\end{theorem}

The nondeterministic state complexity of $\complement{L}$ is bounded by the nondeterministic state complexity of $\complement{L}^\ell$, as the following theorem states.
\begin{theorem}\label{theorem:nsc-complement}
	Let $L\subseteq\alphabet^\ell$, with $\ell>0$, be a block language, such that $\nsc(L)=m$. Then, $\nsc(\complement{L})\in2^{\Omega\left(\sqrt{m}\right)}$.
\end{theorem}

\begin{proof}
	Let $\aut{A}_1$ be a minimal \nfa for $L$ with $m$ states. Based on the results in Lemma~\ref{lemma:nsc-bcomplement-sufficient} and Theorem~\ref{theorem:nsc-bcomplement}, an \nfa~$\aut{A}_2$ for $\complement{L}^\ell$ will have at most $2^{\Omega\left(\sqrt{m}\right)}$ states. Additionally, an \nfa~$\aut{A}_3$ for $\extended{(\complement{L}^\ell)}$ will, in the worst-case, require $\ell$ more states than $\aut{A}_2$, as given in Theorem~\ref{theorem:sc-extension}. As $\complement{L}=\extended{(\complement{L}^\ell)}$, the number of states needed for $\complement{L}$ will be bounded by $2^{\Omega\left(\sqrt{m}\right)}$, as desired, since $m\gg\ell$.
	
	To show that it is tight, recall the language $L_{k, d}$ presented in Section~\ref{sec:bcomplement}, for $k\geq 2$ and $d\geq 2$, and consider the language $\complement{L}_{k,d}$. It is possible to see that every automaton for $\complement{L}_{k,d}$  also requires at least $k^d$ states on the $d$-th rank, just as every automaton for $\complement{L}_{k,d}^{2d}$ does (by Lemma~\ref{lemma:nsc-bcomplement}). Then, Theorem~\ref{theorem:nsc-bcomplement} can be reinvoked to guarantee that $\complement{L}_{k,d}$ requires at least $2^{\Omega\left(\sqrt{m}\right)}$ states.
\end{proof}

\section{Conclusions} \label{sec:conclusions}
In this work, we present the representation of a block language as a binary word, such that each bit indicates whether the corresponding word, according to the lexicographical order, belongs, or not, to the language. Subsequently, we studied the conversion of these representations into minimal machines that recognize the respective language, specifically deterministic finite automata and nondeterministic finite automata, where the latter construction turns out to be a hard problem to solve. Then, we focused on the analysis of the state complexity of block languages. We establish bounds on the maximal size of minimal \nfas for block languages. We also examine the operational state complexity of standard operations applied to block languages. The complexities obtained for operations on block languages are summarized in Table~\ref{tab:cblock}, and one can compare these results with the ones for finite languages summarized in Table~\ref{tab:cfin}. For Boolean operations, the bounds are given using the rank widths and are smaller than the ones for finite languages. For the deterministic state complexity of concatenation and Kleene star, the bounds correspond to special cases of the ones for finite languages. For reversal, the results are analogous to those for finite languages, where they depend on the cost of determinization. However, for block languages, this cost is known to be lower. The remaining results for nondeterministic state complexity meet the values known for finite languages except for the specific operations for block languages (block complement, block language cover, word addition, and word removal).

\begin{table}
	\centering
{	\begin{tabular}{lcccccccc}
		\toprule 
		\multicolumn{7}{c}{Block Languages} \\
		\midrule
		 &
		\multicolumn{1}{c}{sc} & \multicolumn{1}{c}{$|\alphabet|$} & Ref.& \multicolumn{1}{c}{nsc} & \multicolumn{1}{c}{$|\alphabet|$} & Ref. \\
		\midrule
		$L_1\cup L_2$ & $\sum_{i=1}^{\ell-1}(m_in_i+m_i+n_i)+3$ & $3$ &Th.\ref{theorem:sc-union}& $m+n-2$ & $2$& Th.\ref{theorem:nsc-union}\\
		$L_1\cap L_2$ & $\sum_{i=0}^{\ell}m_in_i+1$ & $2$&Th.\ref{theorem:sc-intersection}&   $\sum_{i=0}^{\ell}m_in_i$ & $2$&Th.\ref{theorem:nsc-intersection} \\
		$L_1L_2$ & $m+n-2$ & $1$&Th.\ref{theorem:sc-concatenation} & $m+n-1$ & $1$&Th.\ref{theorem:nsc-concatenation} \\
		$\alphabet^\ell\setminus L$ & $m+\ell-1$ & $2$&Th.\ref{theorem:sc-bcomplement} & $O(2^{\sqrt{m}})$ & $2$ &Th.\ref{theorem:nsc-bcomplement}\\
		$\complement{L}$ & $m$ & $1$ &Th.\ref{theorem:sc-complement} & $O(2^{\sqrt{m}})$ & $2$ &Th.\ref{theorem:nsc-bcomplement}\\
		$L\cup\{w\}$ & $m+\ell-1$ & $2$&Th.\ref{theorem:sc-word-op} & $m+\ell-1$ & $2$&Th.\ref{theorem:nsc-word-op} \\
		$L\setminus\{w\}$ & $m+\ell-1$ & $2$ &Th.\ref{theorem:sc-word-op}& $m+\ell-1$ & $2$ &Th.\ref{theorem:nsc-word-op}\\
		$L^*$ & $m-1$ & $1$ &Th.\ref{theorem:dnsc-kleene}& $m-1$ & $1$ &Th.\ref{theorem:dnsc-kleene}\\
		$L^+$ & $m$ & $1$& Th.\ref{theorem:dnsc-plus}& $m$& $1$ &Th.\ref{theorem:dnsc-plus}\\
		$\reversal{L}$ & $2^{\Theta(\sqrt{m})}$ & $2$ &Th.\ref{theorem:sc-reversal}& $m$ & $1$ &Th.\ref{theorem:nsc-reversal}\\
		$\extended{L}$ & $m+\ell-1$ & $2$ &Th.\ref{theorem:sc-extension}
& $m+\ell$ & $2$ &Th.\ref{theorem:sc-extension}\\
		\bottomrule  
	\end{tabular}}
	\caption{State complexity and nondeterministic state complexity for basic operations on block languages.} 
	\label{tab:cblock}
\end{table}

\bibliographystyle{splncs04}
\bibliography{references}

\end{document}